\documentclass[reqno]{amsart} \usepackage{amscd}
\usepackage{epsf}
\newtheorem{theorem}{Theorem}[section]
\newtheorem{proposition}[theorem]{Proposition}
\newtheorem{lemma}[theorem]{Lemma}

\newtheorem{alphthm}{Theorem}

\theoremstyle{definition}

\theoremstyle{remark} \newtheorem{remark}[theorem]{Remark}
\numberwithin{equation}{section}
\newcommand{\field}[1]{\ensuremath{\mathbb{#1}}}
\newcommand{\CC}{\field{C}}

\newcommand{\HH}{\field{H}}
\newcommand{\PP}{\field{P}}
\newcommand{\RR}{\field{R}}

\newcommand{\ZZ}{\field{Z}}
\newcommand{\complex}[1]{\mathsf{#1}} 
\newcommand{\EEE}{\complex{E}}
\newcommand{\SSS}{\complex{S}}
\newcommand{\BBB}{\complex{B}}
\newcommand{\KKK}{\complex{K}}
\newcommand{\AAA}{\complex{A}}
\newcommand{\CCC}{\complex{C}}

\newcommand{\diff}[1]{\mathcal{#1}} 
\newcommand{\op}[1]{\mathbf{#1}} 
\renewcommand{\d}{\operatorname{d}}
\DeclareMathOperator{\id}{id}
\DeclareMathOperator{\I}{I}
\DeclareMathOperator{\Tot}{Tot}
\DeclareMathOperator{\Hom}{Hom}

\DeclareMathOperator{\Ker}{Ker}
\DeclareMathOperator{\im}{Im}
\DeclareMathOperator{\SL}{SL}
\DeclareMathOperator{\PSL}{PSL}
\newcommand{\del}{\partial}
\newcommand{\delb}{\bar\partial}
\newcommand{\delp}{\partial^\prime}
\newcommand{\delpp}{\partial^{\prime\prime}}
\newcommand{\deltap}{\d}
\newcommand{\deltapp}{\delta}
\newcommand{\eqdef}{\overset{\text{def}}{=}}
\newcommand{\var}{\boldsymbol{\delta}}
\newcommand{\qc}{\mathcal{Q}\mathcal{C}(\Gamma,\tilde\Gamma)}

\begin{document}
\begin{flushright}
\texttt{hep-th/9606163}
\end{flushright}
\title[Generating Functional and Effective
Action]{Generating Functional in CFT and Effective Action
for Two-Dimensional Quantum Gravity on Higher Genus Riemann
Surfaces}
\author{Ettore Aldrovandi} \address{Department of
Mathematics\\ SUNY at Stony Brook\\ Stony Brook, NY
11794-3651\\ USA} \email{ettore@math.sunysb.edu}
\author{Leon A. Takhtajan} \email{leontak@math.sunysb.edu}
\begin{abstract}
We formulate and solve the analog of the universal Conformal
Ward Identity for the stress-energy tensor on a compact
Riemann surface of genus $g>1$, and present a rigorous
invariant formulation of the chiral sector in the induced
two-dimensional gravity on higher genus Riemann
surfaces. Our construction of the action functional uses
various double complexes naturally associated with a Riemann
surface, with computations that are quite similar to descent
calculations in BRST cohomology theory. We also provide an
interpretation of the action functional in terms of the
geometry of different fiber spaces over the Teichm\"{u}ller
space of compact Riemann surfaces of genus $g>1$.
\end{abstract}
\maketitle
\section{Introduction}
Conformal symmetry in two dimensions, according to Belavin,
Polyakov, and Zamolodchikov~\cite{bpz}, is generated by the
holomorphic and anti-holomorphic components $\op{T}(z)$ and
$\bar{\op{T}}(\bar{z})$ of the stress-energy tensor of a
Conformal Field Theory. These components satisfy the
Operator Product Expansions~\cite{bpz,fqs}
\begin{align*}
\op{T}(z)\, \op{T}(w) &\sim \frac{c/2}{(z-w)^4}+\biggl(%
\frac{2}{(z-w)^2}+\frac{1}{z-w}\,\frac{\del}{\del w}
\biggr)\op{T}(w), \\
\bar{\op{T}}(\bar{z})\,\bar{\op{T}}(\bar{w}) &\sim
\frac{c/2}{(\bar{z}-\bar{w})^4}+\biggl(%
\frac{2}{(\bar{z}-\bar{w})^2}+\frac{1}{\bar{z}-\bar{w}}\,
\frac{\del}{\del \bar{w}} \biggr)\bar{\op{T}}(\bar{w}),\\
\op{T}(z)\,\bar{\op{T}}(\bar{w}) &\sim 0,
\end{align*}
where $c$ is the central charge of the CFT and $\sim$ means
``up to the terms that are regular as $z \rightarrow
w$''. These OPE, together with the regularity condition
$\op{T}(z) \sim 1/z^4$ as $\vert z\vert \rightarrow \infty$,
are used to construct Verma modules for the Virasoro algebra
that correspond to the holomorphic and anti-holomorphic
sectors of a CFT. The operator content of the CFT is
specified by the highest weight vectors of the Virasoro
algebra that correspond to the primary fields $\op{O}_l(z,
\bar{z})$ with conformal weights $(h_l, \bar{h}_l)$,
satisfying
\begin{equation*}
\op{T}(z)\, \op{O}_l(w,\bar{w}) \sim \biggl(%
\frac{h_l}{(z-w)^2}+\frac{1}{z-w}\, \frac{\del}{\del
w}\biggr)\, \op{O}_l(w,\bar{w})\, ,
\end{equation*}
and similar OPE with $\bar{\op{T}}(\bar{z})$.

A CFT is determined by the complete set of correlation
functions among the primary fields, which are built up of
conformal blocks: the correlation functions for the
holomorphic sector. The conformal blocks are defined by the
Conformal Ward Identities of BPZ~\cite{bpz}, which follow
from the OPE for the primary fields. Introducing the
generating functional for the $n$-point correlation
functions
\begin{align*}
\exp\{-W[\mu](z_1, \dots, z_n)\} &= \bigl\langle%
\op{O}_1(z_1) \dots \op{O}_n(z_n) \exp\biggl(-\frac {1}{\pi}
\int_{\CC} \mu(z, \bar{z})\,\op{T}(z) \d^2 z
\biggr)\bigr\rangle \\
&\eqdef \langle \op{O}_1(z_1) \cdots
\op{O}_n(z_n)\rangle_{\mu},
\end{align*}
where the integration goes over the complex plane $\CC$ and
$\d^2\! z=\frac{i}{2} \d z \wedge \d\bar{z}=\d x \wedge \d
y,~z=x+iy, \bar{z}=x-iy$, the CWI can be written in the
following ``universal form'' (cf.~\cite{yos, ver})
\begin{equation*}
(\delb-\mu\,\del -2\,\mu_z)\frac{\var W}{\var \mu(z)}=
\frac{c}{12 \pi}\mu_{zzz}+\sum_{l=1}^n \bigl\{h_l \,
\delta_{z}(z-z_l)+\delta(z-z_l)\, \frac{\del W}{\del z_l}
\bigr\},
\end{equation*}
where $\del=\del/\del z$,
$\delb=\del/\del\bar{z}$. Describing the complete solution
of this equation, as well as of its generalization for
higher genus Riemann surfaces, is one of the major problems
of CFT.

This problem remains non-trivial even in the simplest case
of conformal blocks without primary fields, when the
generating functional $W[\mu]$ takes the form
\begin{equation} \label{generating}
\exp\{-W[\mu]\}=\bigl\langle\exp\biggl\{-\frac{1}{\pi}
\int_{\CC} \mu(z, \bar{z})\, \op{T}(z)\,\d^2\! z\biggr\}
\bigr\rangle \eqdef\langle \I \rangle_{\mu}\, .
\end{equation}
It gives the expectation value of the unit operator $\I$ in
the presence of the Schwinger's source term $\mu$, which is
a characteristic feature of all CFT with the same central
charge $c$. The corresponding universal CWI reduces to the
equation
\begin{equation} \label{WI}
(\delb-\mu\,\del-2\,\mu_z)\frac{\var W}{\var \mu(z)}
=\frac{c}{12 \pi} \mu_{zzz}
\end{equation}
for the expectation value of the stress-energy tensor
\begin{equation*}
\langle \op{T}(z) \rangle_{\mu}\eqdef \frac{\var W}{\var
\mu(z)}\, .
\end{equation*}

It is remarkable that the functional $W[\mu]$, for
$\vert\mu\vert<1$, can be determined in closed form and that
it turns out to be the Euclidean version of Polyakov's
action functional for two-dimensional induced quantum
gravity~\cite{pol}.

To see this, let $\mu$ be a Beltrami coefficient on
$\CC$---a bounded function $\mu$ with the property
$\vert\mu\vert <1$---to which one can associate a
self-mapping $f:\CC \rightarrow\CC$ as a unique normalized
(fixing $0,1$ and $\infty$) solution of the Beltrami
equation
\begin{equation*}
f_{\bar{z}}=\mu\, f_z\, .
\end{equation*}
Denote by
\begin{equation*}
T(z)=\{f,z\}=\frac{f_{zzz}}{f_z}-\frac{3}{2}
\biggl(\frac{f_{zz}}{f_z}\biggr)^2
\end{equation*}
the Schwarzian derivative of $f$---``the stress-energy
tensor associated with $f$''. Then (see,
e.g.~\cite{laz,yos}), equation \eqref{WI} is equivalent to
the following Cauchy-Riemann equation
\begin{equation*}
(\delb-\mu \del)\biggl(\bigl(\frac{\var
W}{\var\mu(z)}-\frac{c}{12 \pi} T(z) \bigr)/(f_z)^2\biggr)=0
\end{equation*}
with respect to the complex structure on $\CC$ defined by
the coordinates $\zeta=f(z,\bar{z})$,
$\bar{\zeta}=\overline{f(z,\bar{z})}$.  Using the regularity
of the stress-energy tensor at $\infty$ one gets that
\begin{equation} \label{S}
\frac{\var W}{\var\mu(z)}=\langle \op{T}(z)
\rangle_{\mu}=\frac{c}{12 \pi}T(z) \,.
\end{equation}

This variational equation for determining $W$ was explicitly
solved by Haba~\cite{hab}. Specifically, let $f^{t\mu}$ be
the family of self-mappings of $\CC$ associated to the
Beltrami coefficients $t\mu, \, 0 \leq t \leq 1$. Then
\begin{equation*}
W[\mu]=\frac{c}{12 \pi}\int_{0}^1\d t\int_{\CC}
T^{t\mu}\,\mu\,\d^2\! z
\end{equation*}
solves \eqref{S}. The functional $W$ can be considered as a
WZW type functional since its definition requires an
additional integration over a path in the field space.

Next, consider Polyakov's action functional for
two-dimensional induced quantum gravity in the light-cone
gauge~\cite{pol}, applied to the quasi-conformal map $f$:
\begin{equation}\label{polyakov-action}
S[f]=-\int_{\CC} \frac{f_{zz}}{f_z}\biggl(%
\frac{f_{\bar{z}}}{f_z}\biggr)_{\!z}\d^2\! z\, .
\end{equation}
It has the property
\begin{equation*}
\frac{\var S}{\var \mu(z)}=2\,T(z)=2\{f,z\},
\end{equation*}
so that $c\,S[f]/24 \pi$, considered as a functional of
$\mu=f_{\bar{z}}/f_z$, also solves equation
\eqref{WI}. Therefore, one has the fundamental relation
\begin{equation} \label{ws}
W[\mu]=\frac{c}{24 \pi} S[f],
\end{equation}
which expresses $W$ as a local functional of $f$ and which
can be verified directly. This relation provides the
interpretation (cf.~\cite{yos,bel,pol2}) of two-dimensional
induced gravity in the conformal gauge in terms of a
gravitational WZNW model (and hence in terms of a
Chern-Simons functional as well).

In the present paper we formulate and solve the analog of
the equation \eqref{WI} for the stress-energy tensor on a
compact Riemann surface of genus $g>1$. As in the genus zero
case, it provides an invariant formulation of the chiral
sector in two-dimensional induced gravity on higher genus
Riemann surfaces, a solution to the problem discussed
in~\cite{ver}. From a different point of view, this problem
was also considered in~\cite{z1,z2}.

First, it should be noted that it is trivial to generalize
the genus zero treatment to the case of elliptic
curves---compact Riemann surfaces of genus $1$. Namely, let
$X$ be an elliptic curve realized as the quotient
$L\backslash\CC$ of the complex plane $\CC$ by the action of
a rank $2$ lattice $L$ generated by $1$ and $\tau$, with
$\im\tau>0$. The analog of the equation \eqref{WI} has the
same form, where $\mu$ is now a doubly-periodic function on
$\CC$, while the corresponding normalized solution $f$ of
the Beltrami equation has the property
\begin{equation*}
f(z+1)=f(z)+1\,,\quad f(z+\tau)=f(z)+\tilde{\tau}\,,
\end{equation*}
where $\tilde{\tau}=f(\tau)\,,\;\im\tilde\tau\neq 0$. It
follows that 
\begin{equation*}
f\circ\gamma=\tilde\gamma\circ f \quad\text{for all
$\gamma\in L$,}
\end{equation*}
where $\tilde\gamma\in\tilde{L}$, the rank $2$ lattice in
$\CC$ generated by $1$ and $\tilde\tau$. As a result, the
functional $S[f]$ has the same form as in
\eqref{polyakov-action}, where now the integration goes over
the fundamental parallelogram $\Pi$ of the lattice $L$.

Having thus addressed the genus $1$ case, we start by
formulating equation \eqref{WI}---the same applies to the
universal CWI as well---on a compact Riemann surface $X$ of
genus $g>1$.  In order to do it one needs to use projective
connections on $X$ (see, e.g., \cite{gun} for
details). Namely, recall~\cite{fs} that the stress-energy
tensor $\op{T}$ of a CFT on a Riemann surface is $c/12$
times a projective connection. Therefore the expectation
value
\begin{equation*}
\langle\op{T}(z)\rangle=\frac{c}{12}Q(z),
\end{equation*}
is a holomorphic projective connection on $X$ which depends
on the particular CFT. The difference between two projective
connections on $X$ is a quadratic differential, so that in
order to define the generating functional for the
stress-energy tensor on $X$, one can choose a ``background''
holomorphic projective connection $R$ and set
\begin{equation*} 
\exp\{-W[\mu]\}=\bigl\langle \exp
\biggl\{-\frac{1}{\pi}\int_X \mu(z,\bar{z})\,
(\op{T}(z)-\frac{c}{12}R(z))\d^2\! z\,\biggr\}
\bigr\rangle\,,
\end{equation*}
where $\mu$ is a Beltrami differential on $X$. The analog of
equation \eqref{WI} takes the form~\cite{b,laz}
\begin{equation*}
(\delb-\mu\del-2\mu_z)\frac{\var W}{\var\mu(z)}=\frac{c}{12
\pi}(\mu_{zzz} +2R\mu_z+R_z\mu),
\end{equation*}
where $z$ is a local complex coordinate on $X$, and was used
in~\cite{z1,z2}. As it follows from the definition of $W$,
\begin{equation*}
\frac{\var W}{\var\mu(z)}\bigg|_{\mu=0}=\langle\op{T}(z)
-\frac{c}{12}R(z) \rangle=\frac{c}{12}(Q(z)-R(z))
\end{equation*}
and this expectation value can be set to zero if one chooses
$Q=R$. However, when working with all conformal field
theories on $X$ having the same central charge $c$, it is
preferrable to have a canonical choice of the holomorphic
projective connection $R$. One possibility, which is the
choice we will adopt in this paper, is to use a Fuchsian
projective connection. It is defined by the Fuchsian
uniformization of the Riemann surface $X$, i.e.~by its
realization as a quotient $\Gamma\backslash \HH$ of the
upper half-plane $\HH$ by the action of a strictly
hyperbolic Fuchsian group $\Gamma$ with $2g$ generators. The
upper half plane is isomorphic to the universal cover of
$X$, while $\Gamma$, as an abstract group, is isomorphic to
$\pi_1(X)$, the fundamental group of the surface $X$. Note
that the Fuchsian uniformization of Riemann surfaces plays a
fundamental role in the geometric approach to the
two-dimensional quantum gravity through quantum Liouville
theory (see~\cite{t} and references therein).

The covering $\HH \rightarrow X$ allows to pull-back
geometric objects from $X$ to $\HH$. Since the Fuchsian
projective connection tautologically vanishes on $\HH$, the
stress-energy tensor $\op{T}(z)$ becomes a quadratic
differential for the Fuchsian group $\Gamma$
\begin{equation*}
\op{T} \circ \gamma\, (\gamma^{\prime})^2=\op{T}\quad
\text{for all $\gamma \in\Gamma$,}
\end{equation*}
whereas the source term $\mu$ becomes a Beltrami
differential for $\Gamma$
\begin{equation*}
\mu \circ \gamma\,
\frac{\overline{\gamma^{\prime}}}{\gamma^{\prime}}=\mu\quad
\text{for all $\gamma \in \Gamma$.}
\end{equation*}
The product $\op{T}\mu$ is a $(1,1)$-tensor for $\Gamma$, so
that the integral
\begin{equation*}
\int_{F}\op{T}\mu\, \d z \wedge \d \bar{z}
\end{equation*}
---the natural pairing between quadratic and Beltrami
differentials---is well-defined, i.e.~it does not depend on
the choice of the fundamental domain $F\subset\HH$ of the
Fuchsian group $\Gamma$. As a result, the functional
$W[\mu]$ retains the same form as in formula
\eqref{generating}, where now the integration goes over the
domain $F$, and satisfies the same equation \eqref{WI}, with
$z\in\HH$. It should be noted that the expectation value of
$\langle\op{T}(z)\rangle_{\mu}$ is no longer zero when
$\mu=0$, but rather is $c/12$ times a holomorphic quadratic
differential $q$, which is the pull-back to $\HH$ of the
quadratic differential $Q-R$ on $X$ and characterizes a
particular CFT. Thus, as it was observed in~\cite{z1,z2}, the 
generating functional for
the stress-energy tensor on a higher genus Riemann surface
is no longer a universal feature of all conformal field
theories with the same value of $c$. However, as we shall
show in the paper, one can still find the general solution
of the equation \eqref{WI}.

Next, in order to solve the universal CWI and to define an
action functional for the chiral sector in two-dimensional
induced gravity on $X$, one could first try to extend
Polyakov's functional \eqref{polyakov-action} from $\CC$ to
$X$ by considering the following integral
\begin{equation}\label{int-omega}
\frac{1}{2i}\int_F\omega[f]\, ,
\end{equation}
where
\begin{equation*} 
\omega[f]=\frac{f_{zz}}{f_z}\,
\biggl(\frac{f_{\bar{z}}}{f_z}\biggr)_{\!z}\d
z\wedge\d\bar{z}\,,
\end{equation*} 
which was the correct choice for the genus $1$ case. In
this expression $\mu=f_{\bar{z}}/f_z$ should be a Beltrami
differential for $\Gamma$, which is necessary for an
invariant definition of the generating functional
$W[\mu]$. This imposes strong conditions on the possible
choices of the mapping $f$. It should be noted in the first
place that, contrary to the genus zero case, the
correspondence $f \mapsto \mu(f)=f_{\bar{z}}/f_z$ is no
longer one-to-one. Indeed, the solution of the Beltrami
equation
\begin{equation*}
f_{\bar{z}}=\mu f_z
\end{equation*}
on $\HH$ depends on the extension of the Beltrami
coefficient $\mu$ to the lower half-plane $\overline{\HH}$
of the complex plane $\CC$. There are two canonical choices
compatible with the action of $\Gamma$. In the first case
\begin{equation*}
\mu(\bar{z},z) \eqdef \overline{\mu(z,\bar{z})},\quad z \in
\HH\,,
\end{equation*}
whereas in the second case
\begin{equation*}
\mu(z,\bar{z}) \eqdef 0,\quad z \in \overline{\HH}\,.
\end{equation*}
In both cases, the property of $\mu$ being a Beltrami
differential for $\Gamma$ is equivalent to the following
equivariance property of $f$ (the solution of the Beltrami
equation in $\CC$). There should exist an isomorphism
$\Gamma\ni\gamma\mapsto\tilde{\gamma}\in \tilde{\Gamma}
\subset \PSL(2,\CC)$, such that
\begin{equation} \label{eq}
f \circ \gamma =\tilde{\gamma}\circ f \quad\text{for all
$\gamma \in \Gamma$.}
\end{equation}
In the first case, the restriction of $f$ to $\HH$ yields a
self-mapping of $\HH$ with $\tilde{\Gamma}$ a Fuchsian group
(thus defining a Fuchsian deformation of $\Gamma$), whereas
in the second case $f$ maps $\HH$ onto the interior of a
simple Jordan curve in $\CC$ with $\tilde{\Gamma}$ a
quasi-Fuchsian group (thus defining a quasi-Fuchsian
deformation of $\Gamma$).

However, using the equivariance property of $f$ it is easy
to see that the ``naive'' expression \eqref{int-omega} can
not be considered as a correct choice for the action
functional in higher genus. Indeed, it follows from
\eqref{eq} that:
\begin{enumerate}
\item\label{ein} The density $\omega[f]$ is not a
$(1,1)$-tensor for $\Gamma$, so that the integral
\eqref{int-omega} depends on any particular choice of the
fundamental domain $F$.
\item\label{zwei} The formal variation of \eqref{int-omega}
depends on the values of $\var f$ on the boundary $\del F$
of $F$.
\end{enumerate}

One may try to overcome these difficulties and resolve the
second problem by adding suitable ``correction terms'' to
the functional \eqref{int-omega}; these can be determined by
performing the formal variation of \eqref{int-omega}.
Specifically, all local computations will be the same as in
the genus zero case (see Lemma~\ref{set2}), except that now
\eqref{eq} does not allow to get rid of the boundary terms
in the Stokes formula by setting the variations $\var\mu$ or
$\var f$ to zero on $\del F$. Therefore, besides the local
``bulk'' term, the variation of \eqref{int-omega} will
contain ``total derivative'' terms localized at $\del
F$. This suggests the addition of ``counterterms'', which
depend only on the edges of $F$, such that their variation
cancels the boundary terms coming from the variation of
\eqref{int-omega}. Such counterms can be determined; it
should be noted that a similar, though much simpler
procedure was used in~\cite{zt}, where the Liouville action
functional on the fundamental domain of a Schottky group was
defined. In our case, however, the actual construction goes
one step further: the variation of the edge terms produces
additional quantities localized at the vertices of $\del
F$. In turn, their cancellation requires counterterms that
depend on the vertices of $\del F$, which can be determined
as well.

It turns out that this rather complicated procedure, which
solves problem~\ref{zwei}, can be carried out in a
canonical way using standard tools from homological algebra,
namely various double complexes naturally associated with
the Riemann surface $X$. It is remarkable that at the same
time it solves problem~\ref{ein} as well!

By using the action of the group $\Gamma$ on $\HH$, we
extend the singular chain boundary differential and the
de~Rham differential on \HH\ to act on chains and cochains
for the group homology and cohomology of $\Gamma$. The
corresponding group boundary and coboundary differentials
give rise to two double complexes such that the fundamental
domain $F$ and the density $\omega[f]$ can be extended to
representatives of suitable homology and cohomology classes
$[\Sigma]$ and $[\Omega_f]$ and the pairing between them
becomes $\Gamma$-invariant. Subsequently, we define the
action functional $S[f]$ as the result of such pairing,
i.e. as the evaluation of $[\Omega_f]$ on $[\Sigma]$. Quite
naturally, the actual computation of these representatives
goes exactly like descent calculations, familiar from BRST
cohomology (see, e.g.~\cite{ks}). This is more than a simple
analogy in the following sense. The appropriate tool for
linearizing the action of a discrete group is the group
ring, which leads to the group (co)homology that we are
using for the action of the Fuchsian group $\Gamma$ on
$\HH$. The corresponding concept in the case of a continuous
(Lie) group is the Lie algebra and its (co)homology, which
is used in BRST theory.

The action functional $S[f]$ resulting from this
construction looks as follows. Let $F$ be a canonical
fundamental domain for $\Gamma$ in the form of a closed
non-Euclidean polygon in \HH\ with $4g$ edges. For any
$\gamma \in \Gamma$ and any pair $(\gamma_1, \gamma_2) \in
\Gamma \times \Gamma$, let $\theta_{\gamma}[f]$ and
$\Theta_{\gamma_1,\gamma_2}[f]$ be a $1$-form and a function
on $\HH$ given by the following explicit expressions:
\begin{gather*}
\theta_{\gamma^{-1}}[f] = \log ({\tilde\gamma}^\prime\circ
f)\d\log f_z -\log (f_z\circ\gamma)\d\log\gamma^\prime
-2\frac{\gamma^{\prime\prime}}{\gamma^\prime}\mu\d\bar z\\
\begin{split}
\d\Theta_{\gamma_2^{-1},\gamma_1^{-1}}[f] &=
f^*\bigl(\log\bigl(
\tilde\gamma_1\circ\tilde\gamma_2\bigr)^\prime
\d\log{\tilde\gamma_2}^\prime\bigr)
+\log\gamma_2^\prime\,\d\log
\bigl(\gamma_1\circ\gamma_2\bigr)^\prime\\ &-\frac{1}{2}
f^*\bigl(\d\bigl(\log{\tilde\gamma_2}^\prime\bigr)^2\bigr)
-\frac{1}{2}\d\bigl(\log\gamma_2^\prime\bigr)^2\, ,
\end{split}
\end{gather*}
where $f^*$ denotes the pull-back of differential forms on
$\HH$ by the mapping $f$. Then
\begin{equation}\label{BigAction}
\begin{split}
2iS[f] &= \int_F\omega[f]
-\sum_{i=1}^g\int_{b_i}\theta_{\beta_i}[f]
+\sum_{i=1}^g\int_{a_i}\theta_{\alpha_i}[f] \\ &\quad
+\sum_{i=1}^g \biggl(\Theta_{\alpha_i,\beta_i}[f](a_i(0)) -
\Theta_{\beta_i,\alpha_i}[f](b_i(0)) +
\Theta_{\gamma_i^{-1},\alpha_i\beta_i}[f](b_i(0)) \biggr)\\
&\quad\quad -\sum_{i=1}^{g-1}
\Theta_{\gamma_g^{-1}\dots\gamma_{i+1}^{-1},
\gamma_i^{-1}}[f](b_g(0))\, .
\end{split}
\end{equation}
Here $a_i$ and $b_i$ are the standard cycles on $X$ viewed
as edges of $F$ with initial points $a_i(0)$ and $b_i(0)$,
$\alpha_i$ and $\beta_i$ are the corresponding generators of
the group $\Gamma$, and $\gamma_i$ stands for the commutator
$[\alpha_i,\beta_i] \eqdef \alpha_i
\beta_i\alpha_i^{-1}\beta_i^{-1}$.

Observe that one can formally set $g=1$ in the
representation \eqref{BigAction}, replacing the non-abelian
groups $\Gamma$ and $\tilde\Gamma$ by the lattices $L$ and
$\tilde{L}$, respectively.  Since in this case
$\gamma^{\prime}=\tilde\gamma^{\prime}=1$ identically, the
differential forms $\theta$ and $\d\Theta$ vanish and the
action functional $S[f]$ is given by the bulk term only.

It is also instructive to compare our construction with that
presented in~\cite{z1,z2}. Namely, in~\cite{z1,z2} a
solution of \eqref{WI} was written directly on a higher
genus Riemann surface equipped with additional
algebro-geometric and/or dissection data. Formally, this
solution also features a bulk term derived from the genus
zero Polyakov action plus contributions of lower degree, but
a rather complicated series of prescriptions is involved in
its definition. In our construction, the functional $S[f]$
is written down on the universal cover $\HH$ and it only
depends on the choice of the normalized solution $f$ of the
Beltrami equation on $\HH$. As a result, it enjoys the same
nice variational properties as in the genus zero
case. Specifically, we summarize our main results as
follows.
\begin{alphthm} \label{theorem:a}
The functional $S[f]$ does not depend on either the choice
of the fundamental domain $F$, or the choice of standard
generators for the Fuchsian group $\Gamma$. It has a
geometrical interpretation as a result of the evaluation map
given by the canonical pairing
\begin{equation*}
H^2(X,\CC)\times H_2(X,\ZZ)\longrightarrow \CC
\end{equation*}
where $\omega[f] -\theta[f] -\Theta[f]$ represents an
element in $H^2(X,\CC)$ depending on $f$ and $F$ is
canonically extended to a representative of the fundamental
class of $X$ in $H_2(X,\ZZ)$.
\end{alphthm}

Since the action functional $S[f]$ is independent of all the
choices made, the corresponding variational problem is
well-defined. We shall consider two versions of it,
depending on whether we choose either $\mu$ or $f$, related
through the Beltrami equation, to be the independent
functional variable. In the first case, the independent
variable belongs to the linear space of Beltrami
differentials for $\Gamma$ and the ``source'' Fuchsian group
$\Gamma$ uniquely determines the ``target'' Fuchsian (or
quasi-Fuchsian) group $\tilde\Gamma=f\circ\Gamma\circ
f^{-1}$ through the solution of the Beltrami equation
(``variation with free endpoint'').  In the second case, the
``target'' group $\tilde{\Gamma}$ and the homomorphism
$\Gamma \rightarrow \tilde \Gamma$ are fixed a priori
(``variation with fixed endpoints'') and the independent
variable $f$ is a self-mapping of $\HH$ (or a mapping of
$\HH$ onto the interior of a simple Jordan curve) satisfying
the equivariance property \eqref{eq}. In both cases it is
guaranteed that the boundary terms arising from
\eqref{int-omega} are taken care of by the counterterms in
\eqref{BigAction}, so that we have
\begin{alphthm} \label{theorem:b}
The variation of the action $S[f]$ with respect to $\mu$ or
$f$ is given by the formulas
\begin{equation*}
\var S[f]=2\int_{F}T(z)\,\var\mu(z)\d^2\! z
\end{equation*}
and
\begin{equation*}
\var S[f]=-2\int_{F}\mu_{zzz}\,\frac{\var f}{f_z}\d^2\! z\,,
\end{equation*}
respectively.
\end{alphthm}

Needless to say, the variational derivatives of $S[f]$---the
quantities $T(z)$ and $\mu_{zzz}$---are, respectively,
$(2,0)$ and $(2,1)$-tensors for $\Gamma$ (see
Lemma~\ref{tensors}) and can be therefore pushed down to the
Riemann surface $X \simeq \Gamma\backslash\HH$.

Note that the critical points of the functional $S[f]$,
considered for the mappings $f$ that intertwine a given
Fuchsian group $\Gamma$ and a Fuchsian (or quasi-Fuchsian)
group $\tilde{\Gamma}$, consist of those maps $f$ such that
the corresponding $\mu=f_{\bar{z}}/f_z$ satisfies the
``equation of motion''
\begin{equation} \label{eq-motion}
\mu_{zzz}=0\,.
\end{equation}
For a given pair $\Gamma,\tilde{\Gamma}$, determining the
critical set of $S[f]$ seems to be a very difficult
problem. However, it is rather easy to find the dimension of
the solution space of the equation \eqref{eq-motion} without
imposing any conditions on the target group $\tilde
{\Gamma}=f \circ \Gamma \circ f^{-1}$. We shall show in
section~\ref{action-higher-genus}, using the Riemann-Roch
theorem, that this dimension is actually $4g-3$.

Critical points of the functional $S[f]$ with respect to the
variation with free endpoint satisfy the equation of motion
$T(z)=0$. They are a subset of the previous ``fixed-end''
critical set (cf. Lemma~\ref{anom} and
Proposition~\ref{restatement}). Again, determining this set
seems to be a non simple task.

As in the genus zero case, it follows from
Theorem~\ref{theorem:b} that $c\,S[f]/24\pi$, considered as
a functional of $\mu=f_{\bar{z}}/f_z$, solves equation
\eqref{WI}, and is a solution local in the map $f$. However,
in the higher genus case the correspondence $\mu \mapsto f$
is no longer one-to-one and, at least, there are two
canonical choices for $f$ producing a Fuchsian or a
quasi-Fuchsian deformation of the Fuchsian group
$\Gamma$. Both the functionals $c\,S[f]/24\pi$ corresponding
to these mappings solve equation \eqref{WI}. We shall show
in section~\ref{two-complex-structures} that the difference
of the corresponding stress-energy tensors is a quadratic
differential for $\Gamma$, which is holomorphic with respect
to the complex structure on $X$ determined by the Fuchsian
and the quasi-Fuchsian deformations of $\Gamma$.

As we already mentioned, in genus zero it is possible to
express the solution of \eqref{S} by integrating along a
linear path in the space of Beltrami coefficients. Actually,
as we show in~\ref{homotopy-section}, any path $\mu(t)$ that
connects $\mu$ to $0$ leads to the same functional. In the
higher genus case, we denote by $f^{\mu(t)}$ the
corresponding solutions of the Beltrami equation on $\HH$
producing either a Fuchsian or a quasi-Fuchsian deformation
of $\Gamma$, depending on the given terminal mapping $f$,
and set
\begin{equation*}
T^{t}(z)=\{f^{\mu(t)},z\}\, .
\end{equation*}
According to Lemma~\ref{tensors}, the definition
\begin{equation} \label{WW}
W[\mu] \eqdef \frac{c}{12\pi} \int^1_0\biggl(\int_{X} T^t
\,\dot{\mu}(t)\,\d^2\! z\biggr) \d t\, ,
\end{equation}
where $\dot{\mu}(t)=d\mu(t)/dt$, makes perfect sense since
the integrand in \eqref{WW}, being a product of a Beltrami
and a quadratic differential for $\Gamma$, is a
$(1,1)$-tensor for $\Gamma$. We have

\begin{alphthm}\label{theorem:c} 
\begin{itemize}
\item[(i)] Let $f$ be either a Fuchsian or a quasi-Fuchsian
solution of the Beltrami equation on $\HH$. Then
\begin{equation*}
W[\mu]=\frac{c}{24 \pi}S[f]\,, 
\end{equation*}
so that the functional $W[\mu]$ does not depend on the
choice of the homotopy $\mu(t)$ and
\begin{equation*}
{\var W}=\frac{c}{12\pi}\int_X T(z)\var\mu(z)\d^2z\,.
\end{equation*} 
\item[(ii)] The functional $W[\mu]$ is a holomorphic
functional of $\mu$ in the quasi-Fuchsian case, while in the
Fuchsian case
\begin{equation*}
\frac{\del^2
W[\epsilon\mu]}{\del\epsilon\del\bar\epsilon}\bigg
|_{\epsilon=0}= -\frac{c}{48\pi}
\int_{F}|\mu|^2y^{-2}\d^2z\,,
\end{equation*}
for Bers harmonic Beltrami differentials $\mu$.
\end{itemize} 
\end{alphthm}

It is worth stressing again that $W$, as defined in
\eqref{WW}, is but one possible solution to the universal
CWI on $X$: we have already noted that the solution
corresponding to a given CFT with central charge $c$ may
differ from \eqref{WW} by a term involving a
$\Gamma$-quadratic differential, which is the expectation
value of the stress-energy tensor of that CFT. (Similar
observations about the lack of uniqueness in the solution to
the CWI due to holomorphic quadratic differentials appear
in~\cite{z1,z2}.) Moreover, the fact that in higher genus
the correspondence $\mu\mapsto f$ ceases to be one-to-one
clearly affects the value of \eqref{WW}, which will depend
on the prescription used to solve the Beltrami
equation. These observations lead to the question of what
features of conformal field theories at central charge $c$
are actually conveyed by \eqref{WW}. Since, according to
Theorem~\ref{theorem:c}, the solution of \eqref{WW}
featuring a quasi-Fuchsian deformation depends
holomorphically on $\mu$, it is therefore natural to
conjecture that the corresponding functional $W[\mu]$ (or
$(c/24\pi)S[f]$, through Theorem~\ref{theorem:c}) represents
a universal feature of all conformal field theories with
central charge $c$.

We also observe that \eqref{WW} can be considered as a WZW
type functional, since it is obtained integrating over a
path in the field space. Theorem~\ref{theorem:c} says that
this term has also a local representation in two
dimensions. This parallels the genus zero situation, where
the Polyakov's action in the light cone gauge can be
actually derived from a WZNW model~\cite{alekseev}. (See
also~\cite{yos,yoshida} for the analogous situation in the
conformal gauge.) In that case, one obtains a local
functional in two dimensions as a consequence of the
topological triviality of the WZW term for the group
$\SL_2(\RR)$.

\subsection*{}
The organization of this paper is as follows. In
section~\ref{genus-zero} we present a consistent formulation
of the two-dimensional induced gravity in the conformal
gauge using quasi-conformal (even smooth) mappings of $\CC$
and without using any analytic continuation from the
light-cone gauge or treating $z$ and $\bar{z}$ as
independent variables. There we gather all results, based on
local computations, that will be used in the subsequent
sections. Needless to say, essentially all these results are
known (see~\cite{hab,pol,yos,yoshida}) and we present them
mainly for the convenience of the reader and in order to
make the paper self-contained. We also discuss in detail the
formulation based on the functional $W[\mu]$
from~\cite{hab}, prove that it coincides with the Polyakov's
action functional (which was implicitly contained
in~\cite{yos}) and compute the Hessians of the action
functionals $S[f]$ and $W[\mu]$.

We start section~\ref{algebra} by briefly discussing the genus
$1$ case. Next, we recall the standard concepts
from homological algebra and differential topology that are
needed to treat the case of higher genus Riemann surfaces,
relegating the proofs of some rather technical results to
the appendix. We then present the explicit construction of
the representatives of the fundamental class $[\Sigma]$ and
the cohomology class $[\Omega_f]$ corresponding to the
fundamental domain $F$ and the density $\omega[f]$,
respectively.

In section~\ref{action-higher-genus} we finally define an
analog of the Polyakov's action functional for the Riemann
surface $X$ of genus $g>1$ and prove
Theorems~\ref{theorem:a}, \ref{theorem:b} and
\ref{theorem:c}. We also prove that the solution space of
the equation $\mu_{zzz}=0$ is $4g-3$-dimensional and compute
the Hessians of the action functionals $S[f]$ and $W[\mu]$.

The relation of the constructions presented in
sections~\ref{algebra} and~\ref{action-higher-genus} with
the geometry of various fiber spaces over the Teichm\"uller
space is analyzed in section~\ref{fiber-spaces}. There we
describe $\exp(-W[\mu])$ as a section of a line bundle over
Teichm\"uller space, making contact with previous work on
the subject. In the last subsection we draw our conclusions
and set some directions for future work.

\section{Generating functional and Polyakov's action in
genus zero}\label{genus-zero}
\subsection{} 
Let $f$ be a normalized self-mapping of the complex plane
$\CC$, i.e.~an orientation preserving diffeomorphism of the
Riemann sphere $\PP^1=\CC \cup \{\infty\}$ fixing
$0,1,\infty$. Define a map $f \mapsto \mu=
\mu(f)=f_{\bar{z}}/f_z$, where $\mu$ is a smooth Beltrami
coefficient on $\CC$: a smooth bounded function such that
$\vert \mu \vert<1$. The following basic result of the
theory of quasi-conformal mappings guarantees that the
correspondence $f \mapsto \mu$ is one-to-one and onto.

\begin{proposition} 
Let $\mu \in L^{\infty}(\CC)$ (the Banach space of
measurable functions with finite $\sup$ norm) such that $
\vert \vert \mu \vert \vert_{ \infty}<1$. Then the Beltrami
equation
\begin{equation} \label{bel}
f_{\bar{z}}=\mu f_z
\end{equation}
has a unique solution $f$ fixing $0, 1, \infty$ which is an
orientation preserving quasi-conformal homeomorphism of
$\CC$. The solution is smooth (real-analytic) whenever $\mu$
is smooth (real-analytic).
\end{proposition}

\begin{proof}See~\cite{ahl}.\end{proof}

Let $\omega[f]$ be the following $(1,1)$-form
\begin{equation}\label{omega}
\omega[f]=\frac{f_{zz}}{f_z}\mu_z \d z \wedge \d\bar{z},
\end{equation}
which (see the introduction) we identify as the density of
the Polyakov's action functional. Here and elsewhere it is
understood that $\mu=\mu(f)$. From now on we also assume
that $f(z,\bar{z})-z \rightarrow 0$ as $\vert z\vert
\rightarrow \infty$ in such a way that the $(1,1)$-form
$\omega[f]$ is integrable on $\CC$. (One can simply consider
$\mu$ with finite support; other less restrictive conditions
for the difference $f(z,\bar{z})-z$ can be formulated in
terms of Sobolev spaces.) Define the functional
\begin{equation} \label{action}
S[f]=\frac{1}{2i}\int_{\CC} \omega[f]=-\int_{\CC}
\frac{f_{zz}}{f_z}\mu_z \d^2\!z,
\end{equation}
\begin{remark} The functional $S[f]$ is the Euclidean
version of the Polyakov's action functional for the
two-dimensional quantum gravity in the light-cone gauge~
\cite{pol}. Let us recall that it can be also formally
obtained (cf.~\cite{ver}) as a ``chiral'' version of the
Liouville action
\begin{equation*}
A[\phi]=\frac{1}{2}\int_{\CC}\sqrt{h}\, (h^{ab}\del_a \phi\,
\del_b \phi + \phi\, R_h),
\end{equation*}
(where $x_1=x,~x_2=y$ and $R_h$ is the curvature of the
background metric $h$), in the following way. Consider the
``metric'' $h=(\d z + \mu \d\bar{z}) \otimes
\d\bar{z},~\mu=\mu(f)$ and set $\phi=\log f_z$. Since
$R_h=2\mu_{zz}$, the integrand in $A[\phi]$ is equal to
\begin{equation*}
\phi_z \phi_{\bar{z}}+2
\mu\bigl(-\frac{1}{2}\phi_z^2+\phi_{zz}\bigr)=
-\frac{f_{zz}}{f_z}\,\mu_z +
2\,\biggl(\mu\frac{f_{zz}}{f_z}\biggr)_{\!\!z}\, .
\end{equation*}
\end{remark}

Let $T=\{f,z\}$ be the Schwarzian derivative of the mapping
$f$.  We have the following identity, which could also be
looked at as an ``equation for the trace
anomaly''~\cite{pol,yoshida}.
\begin{lemma} \label{anom}
\begin{equation*}
(\delb-\mu \del-2\mu_z)T=\mu_{zzz}\, .
\end{equation*}
\end{lemma}
\begin{proof} 
A direct computation using the definitions of $\mu$ and of
the Schwarzian derivative.
\end{proof}

\begin{lemma} \label{set} 
The functional $S[f]$ is smooth in the sense that its
variational derivative $\var S/\var \mu(z)$, defined as
\begin{equation*}
\left.\frac{d}{dt}\right|_{t=0}S(\mu +t\,\var
\mu)=\int_{\CC}\frac{\var S}{\var \mu}\, \var \mu \d^2\! z
\end{equation*}
exists and is given by
\begin{equation*}
\frac{\var S}{\var \mu(z)}=2\, T(z).
\end{equation*}
\end{lemma}
\begin{proof} Starting with the formula
\begin{equation} \label{varmu}
\var\mu=\frac{\var f_{\bar{z}}}{f_z}-\mu\frac{\var
f_z}{f_z}\, ,
\end{equation}
that relates the variations of $\mu$ and $f$, we get by a
straightforward computation
\begin{equation} \label{varomega}
\var \omega=\biggl\{\biggl(\frac{\var\!
f_z}{f_z}\biggr)_{\!\!z}
\mu_z+\frac{f_{zz}}{f_z}\var\mu_z\biggr\}\d z \wedge
\d\bar{z}= -2\, T\var\mu \d z\wedge\d\bar{z}-\d\eta,
\end{equation}
where
\begin{equation*}
\eta[f; \var f] =
\biggl(\frac{f_{zz}\var\! f_{\bar{z}}}{f_z^2} 
+\frac{\mu_z \var\! f_z}{f_z}-
\biggl(\frac{f_{zz}}{f_z}\biggr)_z\mu\biggr)\d z 
+\frac{f_{zz}\,\var\! f_z}{f_z^2}\d\bar{z}\, .
\end{equation*}
\end{proof}
\begin{proposition}
The functional $c\,S[f]/24\pi$ is the unique solution of the
universal CWI for the stress-energy tensor.
\end{proposition}
\begin{proof}
It follows immediately from Lemmas~\ref{anom} and \ref{set}
that $cS[f]/24\pi$, considered as a functional of $\mu$,
satisfies the equation \eqref{WI}
\begin{equation*}
(\delb-\mu \del-2\mu_z)\frac{\var W}{\var\mu(z)}=
\frac{c}{12 \pi} \mu_{zzz}\, .
\end{equation*}

To prove uniqueness, consider the difference
\begin{equation*} 
Q[\mu](z)=\biggl(\frac{\var W}{\var\mu(z)}-\frac{c}{24 \pi}
\frac{\var S}{\var \mu(z)}\biggr) (f_z)^{-2}
\end{equation*}
and observe (cf.~\cite{laz,yos}) that it satisfies the
following equation
\begin{equation*}
(\delb-\mu\del)Q[\mu](z)=0,
\end{equation*}
which shows that $Q[\mu](z,\bar{z})$ is holomorphic with
respect to the new complex structure $\zeta=f(z,\bar{z}),
\bar{\zeta}=\overline{f(z,\bar{z})}$ on $\CC$ defined by the
Cauchy-Riemann operator $\delb-\mu\,\del$. Recalling that
$\var W/\var \mu(z)$, as well as $T(z)$, vanish as $\vert
z\vert \rightarrow \infty$ (regularity of the stress-energy
tensor at $\infty$) we conclude that $Q[\mu]$ is an entire
function of $\zeta$ vanishing at $\infty$, so that
$Q[\mu]=0$. Therefore, the functional
\begin{equation*}
\frac{c}{24 \pi}S[f]=-\frac{c}{24
\pi}\int_{\CC}\frac{f_{zz}}{f_z}\mu_z \d^2\!z
\end{equation*}
solves the universal CWI \eqref{WI} on $\PP^1$.
\end{proof}

Next, we determine the variation of $S$ with respect to $f$
and determine the classical equations of motion: the
critical points $\var S[f]=0$ of the functional $S$.

\begin{lemma}\label{set2}
\begin{equation*}
\var S[f]=-2\int_{\CC}(T_{\bar{z}}-\mu\, T_z -2\,\mu_z\,
T)\,\frac{\var f}{f_z}\, \d^2\! z = -2\int_{\CC} \mu_{zzz}\,
\frac{\var f}{f_z}\, \d^2\! z\, ,
\end{equation*}
so that the classical equation of motion is
\begin{equation*}
\mu_{zzz}=0\, .
\end{equation*}
\end{lemma}
\begin{proof} It follows from the identity 
\begin{equation*}
T\,\var\mu \d z \wedge \d\bar{z} = (-T_{\bar{z}}+\mu\,
T_z+2\mu_z T)\, \frac{\var f}{f_z}-\d\eta^{\prime},
\end{equation*} 
where
\begin{equation*}
\eta^{\prime}=T\frac{\var\! f}{f_z}\d z+ \mu\,T\frac{\var\!
f} {f_z}\d\bar{z}\, ,
\end{equation*} 
and from Lemma~\ref{anom}.
\end{proof}

\subsection{}\label{homotopy-section}
Let $\mu(t)$, $0 \leq t \leq 1$, be the path in the space of
Beltrami coefficients connecting $0$ with the given Beltrami
coefficient $\mu$. It gives rise to a homotopy
$f^t=f^{\mu(t)}$, $f^0=\id$, $f^1=f$ that consists of
normalized quasi-conformal mappings satisfying the Beltrami
equation
\begin{equation*}
f^t_{\bar{z}}=\mu(t) f^t_z\, .
\end{equation*} 
Denoting the corresponding Schwarzians as
$T^t(z)=\{f^t,z\}$, so that $T^0=0$ and $T^1=T$, we have the
following useful variational formulas.
\begin{lemma} \label{formulas} 
\begin{align}
\mu(t)_{zzz} &= (\delb-\mu(t)\,\del-2\,\mu(t)_z)(T^t),
\tag{i}\label{i}\\ \var T^t
&=(\del^3+2\,T^t\,\del+T^t_z)(u^t), \tag{ii}\label{ii}\\
\var\mu(t) &= (\delb-\mu(t)\,\del+\mu(t)_z)(u^t),
\tag{iii}\label{iii}
\end{align}
where $u^t=\var f^t/f^t_z$.
\end{lemma}
\begin{proof}
Equation \eqref{i} is just a restatement of
Lemma~\ref{anom}, applied to the map $f^t$. The variational
formula \eqref{ii} is verified by a straightforward (though
lengthy) computation using $T=\{f,z\}$ and the definition of
the Schwarzian derivative. Finally, equation \eqref{iii}
follows from the variational formula \eqref{varmu}, written
as
\begin{equation*}
\var\mu=(\delb-\mu\,\del+\mu_z)\biggl(\frac{\var\!
f}{f_z}\biggr)
\end{equation*}
and specialized to the map $f^t$.
\end{proof}

As it follows from Lemma~\ref{formulas}, the differential
operators
\begin{gather*}
\diff{T}=\del^3 + 2\,T\, \del +T_z\\ \intertext{and}
\diff{M}=\delb-\mu\,\del+\mu_z.
\end{gather*}
play a fundamental role in the variational theory. In
particular, the third-order differential operator $\diff{T}$
appears in many other different areas as well. It serves as
a Jacobi operator for the second Poisson structure for the
KdV equation~\cite{magri} that is given by the Virasoro
algebra and it plays an important role in Eichler cohomology
on Riemann surfaces~\cite{gun}. The operator $\diff{T}$ is
skew-symmetric, $\diff{T}^{\tau}=-\diff{T}$, with respect to
the inner product given by
\begin{equation} \label{inner}
(u,v)=\int_{\CC} u\, v\, \d^2z\, ,
\end{equation}
whereas $\diff{M}^{\tau}=-\diff{D}$, where $\diff{D} \eqdef
\delb-\mu\del- 2\mu_z$. However, we have the following
result.
\begin{lemma} \label{skew}
The operator $\diff{T}\diff{M}$ is symmetric.
\end{lemma}
\begin{proof}
It reduces to the verification of the identity $(\diff{T}
\diff{M})^{\tau} =\diff{D}\diff{T}$, or
\begin{equation*}
(\del^3+2\,T\,\del+T_z)(\delb-\mu\,\del+\mu_z)
=(\delb-\mu\,\del+2\mu_z)(\del^3 +2\,T\,\del+T_z),
\end{equation*}
which immediately follows from Lemma~\ref{anom} and
$T=\{f,z\}$.
\end{proof}

Now, let us introduce the functional
\begin{equation} \label{W}
W[\mu]=\frac{c}{12 \pi}\int^1_0\int_{\CC} T^t
\,\dot{\mu}(t)\,\d^2\! z\, \d t,
\end{equation}
where the dot stands for $d/dt$. A priori it may depend on
the choice of the homotopy $\mu(t)$. The following result
shows that the variational derivative of $W$ with respect to
$\mu=\mu(1)$ does not depend on $\mu(t)$.
\begin{lemma} \label{varw} 
\begin{equation*}
\frac{\var W}{\var\mu(z)}=\frac{c}{12 \pi}T(z)\, .
\end{equation*}
\end{lemma}
\begin{proof}
Writing $\var (T^t\dot{\mu}(t))=\var T^t \dot{\mu}(t)+T^t
\var \dot{\mu}(t)$ and using \eqref{ii} in Lemma
\ref{formulas}, together with the relation
\begin{equation} \label{mdot}
\dot{\mu}(t)={\mathcal M}^t(v^t),
\end{equation}
(where $v^t=\dot{f}^t/f^t_z$) which follows from formula
\eqref{iii} of Lemma~\ref{formulas} applied to $\var=d/dt$,
we get
\begin{equation*}
\var T^t\dot{\mu (t)}={\mathcal T}^t(u^t){\mathcal
M}^t(v^t)\, . 
\end{equation*}
Using Lemma~\ref{skew}, equations \eqref{mdot}, \eqref{iii}
and the equation
\begin{equation*}
\dot{T}^t={\mathcal T}^t(v^t)\, ,
\end{equation*}
which follows from formula \eqref{ii} of Lemma
\ref{formulas} applied to $\var=d/dt$, we obtain
\begin{equation*}
\begin{split}
\int_{\CC}\var T^t \dot{\mu}(t)\,\d^2z
&=(\diff{T}^t(u^t),\diff{M}^t
(v^t))=-(u^t,\diff{T}^t\diff{M}^t (v^t)) \\
&=(u^t,(\diff{M}^t)^{\tau}\diff{T}^t(v^t))=(\diff{M}^t(u^t),
\diff{T}^t(v^t)) \\ &= \int_{\CC}
\var\mu(t)\dot{T}^t\,\d^2z\,.
\end{split}
\end{equation*}
Substituting this into the expression for $\var W$, we get
\begin{equation*}
\int_0^1(\dot{T}^t \var \mu(t)+T^t \var \dot{\mu}(t))\, \d
t=\left.T^t \var \mu(t) \right|^{t=1}_{t=0}=T \var \mu,
\end{equation*}
which completes the proof.
\end{proof}

Moreover, as the next result shows, the functional $W$ is
actually independent of the choice of the path $\mu(t)$
connecting the points $0$ and $\mu$ in the space of Beltrami
coefficients.
\begin{proposition}
\begin{equation*}
W[\mu]=\frac{c}{24 \pi}S[f]\, ,
\end{equation*}
where $f$ and $\mu$ are related through
$\mu=f_{\bar{z}}/f_z$.
\end{proposition}
\begin{proof}
It is essentially the computation in Lemma~\ref{set}, done
in the reverse order. Namely, considering the families
$\mu(t)$ and $f^{\mu(t)}$ and using the formula
\eqref{varomega} for the case $\var=d/dt$, we get
\begin{equation*}
2\, T^t\dot{\mu}(t) \d z \wedge\d\bar{z}
=\frac{d}{dt}\biggl(\frac{f^{t}_{zz}}{f^{t}_z} \mu(t)_z
\d{z} \wedge\d\bar{z}\biggr)+\d\eta[f^t; \dot{f}^t]\, ,
\end{equation*}
which after integrating over $\CC \times [0,1]$ yields the
result.
\end{proof}
\subsection{}\label{second-variation}
Here we compute the Hessian of the functional $S[f]$,
i.e.~its second variation with respect to $f$, evaluated at
the critical point.  Let $\var_1f$ and $\var_2f$ be two
variations of $f$, defined through the two-parameter family
$f_{s,t}$ with $f_{0,0}=f$ as
\begin{equation*}
\var_1f=\left.\frac{\del f_{s,t}}{\del s}\right|_{s=t=0}\,
,\quad \var_2f=\left.\frac{\del f_{s,t}}{\del
t}\right|_{s=t=0}\, .
\end{equation*}
The second variation of $S[f]$ is
\begin{equation*}
\var^2S[f]=\left.\frac{d^2}{ds\,
dt}S[f_{s,t}]\right|_{s=t=0}\, ,
\end{equation*}
and it can be computed using the first variation of $S[f]$
from Lemma~\ref{set2}
\begin{equation*}
\var_1S[f]= - 2\int_\CC\mu_{zzz}\,\frac{\var_1f}{f_z}
\,\d^2\!z
\end{equation*}
by evaluating $\var_2(\mu_{zzz}[f])$. As it follows from
Lemma~\ref{formulas},
\begin{equation}\label{second:1}
\var_2\bigl(\mu_{zzz}[f]\bigr)
=\bigl(\del^3\circ\diff{M}\bigr)
\biggl(\frac{\var_2f}{f_z}\biggr)\, ,
\end{equation}
so that
\begin{equation}\label{second-variation:1}
\var^2\, S[f](\var_1f,\var_2f) =-2\int_\CC
\frac{\var_1f}{f_z}\bigl(\del^3\circ\diff{M}\bigr)
\biggl(\frac{\var_2f}{f_z}\biggr)\, \d^2\!z\,.
\end{equation}

The Hessian is symmetric, so that the right hand side of
\eqref{second-variation:1} should be a symmetric bilinear
form in $\var_1f,\var_2f$ whenever $\mu_{zzz}=0$.  This can
be verified directly, as we have

\begin{lemma}
The operator $\del^3\circ\diff{M}$ for $\mu_{zzz}=0$ is
symmetric with respect to the bilinear form \eqref{inner}.
\end{lemma}
\begin{proof}
Using $(\del^3)^\tau=-\del^3$ we have
\begin{equation*}
\bigl(\del^3\circ\diff{M}\bigr)^\tau= \diff{D}\circ\del^3\,,
\end{equation*}
where $\diff{D}=\delb -\mu\,\del -2\,\mu_z$, and it is
straightforward to verify the following identity when
$\mu_{zzz}=0$:
\begin{equation*}
\del^3\circ\diff{M}=\diff{D}\circ\del^3\,.
\end{equation*}
\end{proof}

Similarly, one can compute the Hessian of the functional
$W[\mu]$. We have
\begin{lemma}
\begin{equation*}
\var^2\, W[\mu](\var_1\mu,\var_2\mu)=\frac{c}{12\pi}\int_\CC
\var_1\mu\bigl(\del^3\circ\mathcal{M}^{-1}\bigr)(\var_2\mu)\,
\d^2\!z.
\end{equation*}
\end{lemma}
\begin{remark}
Since
\begin{equation} \label{delbar}
\mathcal{M}\biggl(\frac{u\circ
f}{f_z}\biggr)=\frac{\overline{f_z}}{f_z}\,
(1-|\mu|^2)\,(\delb u)\circ f\,,
\end{equation}
the operator $\mathcal{M}$ is invertible on the subspace of
smooth functions on $\CC$ vanishing at $\infty$.
\end{remark}
\section{Algebraic and topological
constructions}\label{algebra}
\subsection{}\label{genus1}
Here we consider the genus $1$ case. Let $X$ be an elliptic
curve, i.e.~a compact Riemann surface of genus $1$,
realized as the quotient $X\cong L\backslash\CC$, where $L$
is a rank $2$ lattice in $\CC$, generated by the
translations $\alpha(z)=z+1$ and $\beta(z)=z+\tau$, where
$\im\tau>0$. Let $\mu$ be a Beltrami coefficient for $L$,
i.e.~ a $||\mu||_{\infty}<1$ function on $\CC$ satisfying
\begin{equation*}
\mu\circ\gamma=\mu\quad \text{for all $\gamma\in L$,}
\end{equation*}
and let $f=f^{\mu}$ be the normalized (fixing $0,1,\infty$)
solution of the Beltrami equation on $\CC$
\begin{equation*}
f_{\bar{z}}=\mu f_z\,.
\end{equation*}
It is easy to see that $f\circ L=\tilde{L}\circ f$, where
$\tilde{L}$ is the rank $2$ lattice in $\CC$ generated by
$1$ and $\tilde{\tau}=f(\tau)$.  Indeed, $\tilde\gamma=f
\circ\gamma\circ f^{-1}$ is a parabolic element in
$\PSL(2,\CC)$ fixing $\infty$, i.e.~a translation $z\mapsto
z+h$, and it follows from the normalization that
$f(z+1)=f(z)+1$. Therefore the $(1,1)$-form $\omega[f]$ on
$\CC$ is well-defined on $X$ so that the action functional
takes the form
\begin{equation*}
S[f]=\frac{1}{2i}\int_{\Pi}\omega[f]\,,
\end{equation*}
where $\Pi$ is the fundamental parallelogram for the lattice
$L$. 

\subsection{}\label{marking}
Here we consider the higher genus case and construct double
complexes that extend the singular chain and the de~Rham
complexes on \HH\ . We extend the fundamental domain $F$ for
$\Gamma$ and the $(1,1)$-form $\omega[f]$ on $\HH$ to
representatives of the homology and cohomology classes
$[\Sigma]$ and $[\Omega_f]$ for these double complexes.
\subsubsection{}
\begin{figure}
\centering 
\epsfxsize=.30\linewidth
\epsffile{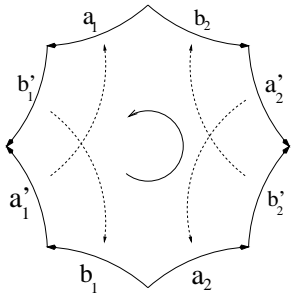}
\caption{Conventions for the fundamental domain $F$}
\label{fig:conventions}
\end{figure}

Let $X\cong\Gamma\backslash\HH$ be a compact Riemann surface
of genus $g>1$, realized as the quotient of the upper
half-plane \HH\ by the action of a strictly hyperbolic
Fuchsian group $\Gamma$.  Recall that the group $\Gamma$ is
called marked if there is a chosen system, up to inner
automorphism, of $2g$ free generators $\alpha_1,\dots
,\alpha_g,\beta_1,\dots , \beta_g$ satisfying the single
relation
\begin{equation} \label{relation} 
[\alpha_1,\beta_1] \cdots [\alpha_g,\beta_g]=1\,,
\end{equation}
where $[\alpha_i,\beta_i] \eqdef
\alpha_i\beta_i\alpha_i^{-1}\beta_i^{-1}$ and $1$ is the
unit element in $\Gamma$. For every choice of the marking
there is a standard choice of a fundamental domain
$F\subset\HH$ for $\Gamma$ as a closed non-Euclidean polygon
with $4g$ edges, pairwise identified by suitable group
elements. We will use the following normalization (see,
e.g.,~\cite{katlok} and Figure~\ref{fig:conventions}). 
The edges of $F$ are labelled by $a_i, a_i^\prime, b_i,
b_i^\prime$ and $\alpha_i (a_i^\prime )=a_i,~ \beta_i
(b_i^\prime ) = b_i$ for all $i=1, \ldots, g$; the
orientation of the edges is chosen so that $\del F =
\sum_{i=1}^g (a_i + b_i^\prime - a_i^\prime - b_i)$. Also we
set $\del a_i=a_i(1)-a_i(0)$ and $\del b_i = b_i(1) -
b_i(0)$, where the label ``1'' represents the end point and
the label ``0'' the initial point with respect to the edge's
orientation.  One has the following relations between the
vertices of $F$ and the generators: $a_i(0)=b_{i+1}(0)$,
$\alpha^{-1}_i(a_i(0))=b_i(1)$,
$\beta^{-1}_i(b_i(0))=a_i(1)$ and
$[\alpha_i,\beta_i](b_i(0))=b_{i-1}(0)$, where, in
accordance with \eqref{relation}, $b_0(0)=b_g(0)$.
%
\subsubsection{}\label{quasi-conformal}
Let $\mu$ be a Beltrami differential for the Fuchsian group
$\Gamma$, i.e.~a bounded ($L^{\infty}(\HH)$) function on
\HH\ satisfying
\begin{equation*}
\mu\circ\gamma\,
\frac{\overline{\gamma^\prime}}{\gamma^\prime} =\mu
\quad\text{for all $\gamma \in \Gamma$.}
\end{equation*}
In addition, it is called a Beltrami coefficient for
$\Gamma$ when $\vert\vert \mu \vert \vert_{\infty} <
1$. Denote by $f=f^{\mu}$ the normalized (fixing $0,1$ and
$\infty$) solution of the Beltrami equation on $\HH$
\begin{equation*}
f_{\bar{z}}=\mu f_z\,.
\end{equation*}
As it was already explained in the introduction, we consider
$f$ to be either a self-mapping of \HH\,, or a mapping of
\HH\ onto the interior of a simple Jordan curve in $\CC$,
uniquely determined by $\mu$. These two choices can be
realized by considering the Beltrami equation on the whole
complex plane $\CC$: in the former case the Beltrami
coefficient $\mu$ is extended to the lower half-plane
$\overline{\HH}$ by reflecting it through the real line
$\RR$, while in the latter $\mu$ is extended by zero in
$\overline{\HH}$. In both cases there exists $\tilde\Gamma
\subset \PSL(2,\CC)$, isomorphic to $\Gamma$ as an abstract
group and such that $f$ intertwines between $\Gamma$ and
$\tilde{\Gamma}$
\begin{equation*}
f \circ \gamma=\tilde{\gamma} \circ f\quad \text{for all
$\gamma \in \Gamma$,}
\end{equation*}
which actually defines the isomorphism $\gamma \mapsto
\tilde{\gamma}$. In the first case we have that
$\tilde\Gamma \subset \PSL(2,\RR)$ and it is in fact a
Fuchsian group, a Fuchsian deformation of $\Gamma$.  In the
second case $\tilde\Gamma$ is a so-called quasi-Fuchsian
group, a special case of a Kleinian group. Its domain of
discontinuity has two invariant components, the interior and
the exterior of a simple Jordan curve in $\CC$, which is the
image of the real line $\RR$ under the mapping $f$ and is a
limit set for $\tilde\Gamma$. These mappings, introduced and
studied by Ahlfors and Bers, play a fundamental role in
Teichm\"{u}ller theory (see, e.g.~\cite{gar}).
\subsubsection{}\label{homology}
Let $\SSS_\bullet\equiv S_\bullet(X_0)$ be the standard
singular chain complex of $\HH$ with the differential
$\delp.$ (From now on, we will denote the singular chain
differential by $\delp$, as the symbol $\del$ will be
reserved for the total differential in a double complex, to
be introduced below.) The group $\Gamma$ acts on $\HH$ and
induces a left action on $\SSS_\bullet$ by translating the
chains, hence $\SSS_\bullet$ becomes a complex of
$\Gamma$-modules. Since the action of $\Gamma$ on \HH\ is
proper, $\SSS_\bullet$ is a complex of left \emph{free}
$\ZZ\Gamma$-modules~\cite{mac-lane}, where $\ZZ\Gamma$ is
the integral group ring of $\Gamma$: the set of finite
combinations $\sum_{\gamma\in\Gamma} n_\gamma\gamma$ with
coefficients $n_\gamma\in\ZZ$.

Let $\BBB_\bullet\equiv B_\bullet(\ZZ\Gamma)$ be the
canonical ``bar'' resolution complex for $\Gamma$, with
differential $\delpp .$ Each $B_n(\ZZ\Gamma)$ is a free left
$\Gamma$-module on generators $[\gamma_1|\dots |\gamma_n]$,
with the differential $\delpp :\BBB_n\rightarrow \BBB_{n-1}$
given by
\begin{equation*}
\begin{split}
\delpp [\gamma_1|\dots|\gamma_n] &= \gamma_1\,
[\gamma_2|\dots |\gamma_n] +
\sum_{i=1}^{n-1}(-1)^i [\gamma_1|\dots
|\gamma_i\,\gamma_{i+1}|\dots |\gamma_n]\\ &\qquad
+(-1)^n[\gamma_1|\dots|\gamma_{n-1}]\, 
\end{split}
\end{equation*}
for $n>1$ and by
\begin{equation*}
\delpp [\gamma]=\gamma [\;]-[\;]
\end{equation*}
for $n=1$. Here $[\gamma_1|\dots |\gamma_n]$ is defined to
be zero if any of the group elements inside $[ \ldots ]$
equals the unit element $1$ in $\Gamma$. $B_0(\ZZ\Gamma)$ is
a $\ZZ\Gamma$-module on one generator $[\;]$, and can be
identified with $\ZZ\Gamma$ under the isomorphism that sends
$[\;]$ to $1$.

Next, consider the double complex $\KKK_{\bullet,\bullet}=
\SSS_\bullet\otimes_{\ZZ\Gamma}\BBB_\bullet$. The associated
total simple complex $\Tot\KKK$ is equipped with the total
differential $\del=\delp + (-1)^p\delpp$ on $\KKK_{p,q}.$
For the sake of future reference, we observe that
$\SSS_\bullet$ is identified with
$\SSS_\bullet\otimes_{\ZZ\Gamma}\BBB_0$ under the
correspondence $c\mapsto c\otimes[\;]$.
\begin{remark}
Since $\SSS_\bullet$ and $\BBB_\bullet$ are both complexes
of left $\Gamma$-modules, in order to define their tensor
product over $\ZZ\Gamma$ we need to endow each $\SSS_n$ with
a right $\Gamma$-module structure. This is done in the
standard fashion by setting $c\cdot\gamma \eqdef
\gamma^{-1}(c).$ As a result $\SSS\otimes_{\ZZ\Gamma}\BBB =
\left(\SSS\otimes_{\ZZ}\BBB\right)_\Gamma$, so that the
tensor product over integral group ring of $\Gamma$ can be
obtained as the set of $\Gamma$-invariants in the usual
tensor product (over $\ZZ$) as abelian groups~\cite{brown}.
\end{remark}
The application of standard spectral sequence machinery,
together with the trivial fact that $\HH$ is acyclic, leads
to the following lemma, whose formal proof immediately
follows, for example, from~\cite{mac-lane}, Theorem XI.7.1
and Corollary XI.7.2.
\begin{lemma}\label{lemma-homology}
There are isomorphisms
\begin{equation*}
H_\bullet(X,\ZZ)\cong H_\bullet(\Gamma,\ZZ)\cong
H_\bullet(\Tot\KKK_{\bullet,\bullet})\,,
\end{equation*}
where the three homologies are the singular homology of $X$,
the group homology of $\Gamma$ and the homology of the
complex $\Tot\KKK_{\bullet,\bullet}$ with respect to the
total differential $\del$.
\end{lemma}

We will use this lemma in the construction of the explicit
cycle $\Sigma$ in $\Tot\KKK$ that extends the fundamental
domain $F$. For the convenience of the reader we present a
simple minded proof of Lemma~\ref{lemma-homology} in
Appendix~\ref{app:a}.
\subsubsection{} 
We now turn to constructions dual to those
in~\ref{homology}. Denote by $\AAA^\bullet\equiv
A^\bullet_\CC(X_0)$ the complexified de Rham complex on
$\HH$. Each $\AAA^n$ is a left $\Gamma$-module with the
pull-back action of $\Gamma$, i.e.~$\gamma\cdot\phi \eqdef
(\gamma^{-1})^*\phi$ for $\phi\in\AAA^\bullet$ and for all
$\gamma\in\Gamma$.  Consider the double complex
$\CCC^{p,q}=\Hom (\BBB_q,\AAA^p)$ with differentials
$\deltap$, the usual de~Rham differential, and
$\deltapp=(\delpp)^*$, the group coboundary. Specifically,
for $\phi\in\CCC^{p,q}$
\begin{equation*}
\begin{split}
(\deltapp\phi)_{\gamma_1,\dots,\gamma_{q+1}} &=
\gamma_1\cdot\phi_{\gamma_2,\dots,\gamma_{q+1}} +
\sum_{i=1}^q\, (-1)^i\phi_{\gamma_1,\dots,\gamma_i
\gamma_{i+1}\dots,\gamma_{q+1}}\\ &\qquad +
(-1)^{q+1}\phi_{\gamma_1,\dots,\gamma_q}\, .
\end{split}
\end{equation*}
As usual, the total differential on $\CCC^{p,q}$ is $D=
\deltap + (-1)^p\deltapp$. Either by dualizing
Lemma~\ref{lemma-homology} or working out the spectral
sequences resulting from $\CCC$, we obtain the
\begin{lemma}\label{lemma-cohomology}
There are isomorphisms
\begin{equation*}
H^\bullet (X,\CC)\cong H^\bullet (\Gamma ,\CC) \cong
H^\bullet (\Tot\CCC^{\bullet,\bullet}) \, ,
\end{equation*}
where the the three cohomologies are the de Rham cohomology
of $X$, the group cohomology of $\Gamma$ and the cohomology
of the complex $\Tot\CCC^{\bullet,\bullet}$ with respect to
the total differential $D$.
\end{lemma}

As for Lemma~\ref{lemma-homology}, a simpler proof can also
be found in Appendix~\ref{app:a}.

Finally, there exists a natural pairing between $\CCC^{p,q}$
and $\KKK_{p,q}$ which assigns to the pair $(\phi ,\,
c\otimes[\gamma_1|\dots|\gamma_q])$ the evaluation of the
form $\phi_{\gamma_1, \ldots, \gamma_q}$ over a cycle $c$
\begin{equation}\label{pairing}
\langle \phi,c\otimes[\gamma_1|\dots|\gamma_q]\rangle =
\int_c\phi_{\gamma_1,\dots,\gamma_q}\, .
\end{equation}
By the very construction of the double complexes
$\CCC^{\bullet,\bullet}$ and $\KKK_{\bullet,\bullet}$, the
total differentials $D$ and $\del$ are transpose to each
other
\begin{equation} \label{dual}
\langle D \Phi,C\,\rangle =\langle\,\Phi,\del C \rangle
\end{equation}
for all $\Phi \in \CCC^{\bullet,\bullet},~C \in
\KKK_{\bullet,\bullet}$.  Therefore the pairing
\eqref{pairing} descends to the corresponding homology and
cohomology groups and is non degenerate. It defines a
pairing between $H^{\bullet}(\Tot\CCC^{\bullet,\bullet})$
and $H_{\bullet}(\Tot\KKK_{\bullet,\bullet})$ which we
continue to denote by $\langle\;,\;\rangle$.
\subsection{}
Here we compute explicit representatives $\Sigma$ and
$\Omega_f$, for the fundamental class of the surface $X$ and
a degree two cohomology class on $X$ that extend the
fundamental domain $F$ and the $2$-form $\omega[f]$,
respectively.

\subsubsection{Homology computations} 
\label{homology-computations}
Fix the marking of $\Gamma$ and choose a fundamental domain
$F$ as in \ref{marking}. We start by the observation that
$F\cong F\otimes [\;]\in\KKK_{2,0}.$ Furthermore, obviously
$\delpp F =0$, and
\begin{equation*}
\begin{split}
\delp F &= \sum_{i=1}^g (b_i^\prime - b_i -a_i^\prime
+a_i)\\ &= \sum_{i=1}^g (\beta_i^{-1}(b_i) - b_i
-\alpha_i^{-1}(a_i) + a_i)\, ,
\end{split}
\end{equation*}
which we can rewrite as $\delp F = \delpp L$ where
$L\in\KKK_{1,1}$ is given by
\begin{equation}
L = \sum_{i=1}^g (b_i\otimes [\beta_i] - a_i\otimes
[\alpha_i])\, .
\end{equation}
This follows from $\gamma^{-1}(c) - c =
c\cdot\gamma-c=c\otimes\gamma[\;] - c\otimes [\;] =
c\otimes\delpp[\gamma]$ for any singular chain $c$ and any
$\gamma\in\Gamma$.

Let us now compute $\delp L$. There exists $V\in\KKK_{0,2}$
such that $\delp L = \delpp V$; its explicit expression is
given by
\begin{equation}
\begin{split}
V &= \sum_{i=1}^g\left(a_i(0)\otimes [\alpha_i|\beta_i] -
b_i(0)\otimes [\beta_i|\alpha_i] + b_i(0)\otimes
[\gamma_i^{-1}|\alpha_i\beta_i] \right)\\
&\quad-\sum_{i=1}^{g-1}b_g(0)\otimes
[\gamma_g^{-1}\dots\gamma_{i+1}^{-1}|\gamma_i^{-1}]\, ,
\end{split}
\end{equation}
where $[\alpha_i,\beta_i]=\gamma_i$. Indeed, a
straightforward computation, using the relations between
generators and vertices, yields
\begin{equation*}
\delp L=\delpp V - b_g(0)\otimes
[\gamma_g^{-1}\dots\gamma_1^{-1}]\, ,
\end{equation*}
and the second term in the RHS vanishes by virtue of
\eqref{relation}, since $[1]=0$.

From the relations $\delp F = \delpp L$ and $\delp L =
\delpp V$ it follows immediately that the element
$\Sigma=F+L-V$ of total degree two is a cycle in
$\Tot\,\KKK$, that is
\begin{equation*}
\del (F+L-V)=0\,.
\end{equation*}
Thus we have the
\begin{proposition}\label{fund-class}
The cycle $\Sigma\in (\Tot\,\KKK)_2$ represents the
fundamental class of the surface in $H_2(X,\ZZ)$.
\end{proposition}
\begin{proof}
This follows immediately from Lemma~\ref{lemma-homology},
provided the class $[\Sigma]$ is not zero, but this is not
the case, since the cycle $\Sigma$ is a ``ladder'' starting
from the fundamental domain $F$. It follows from the
arguments in Appendix~\ref{app:a} that the latter in fact
maps under $\SSS_2\ni F\mapsto F\otimes 1\in
\SSS_2\otimes_{\ZZ\Gamma}\ZZ\cong S_2(X)$ to a
representative of the fundamental class.
\end{proof}
\begin{remark}
The existence of the elements $L$ and $V$ can be guaranteed
a priori by the methods of Appendix~\ref{app:a}, using the
fact that $\Gamma$ has no cohomology except in degree zero.
\end{remark}

As it follows from Proposition~\ref{fund-class}, the
homology class $[\Sigma]$ is independent of the marking of
the Fuchsian group $\Gamma$ and of the choice of the
fundamental domain $F$, whereas its representative $\Sigma$
is not. Since this independence is a key issue in defining
the action functional for the higher genus case, we will
show explicitly that different choices lead to homologous
$\Sigma$. Essentially, these choices are the following.
\begin{itemize}
\item Within the same marking choose another set of
canonical generators $\alpha_i^{\prime},\beta_i^{\prime}$ by
conjugating $\alpha_i,\beta_i$ with $\gamma \in \Gamma$ so
that $F^{\prime}=\gamma F$ for the corresponding fundamental
domains.
\item Within the same marking make a different choice of the
fundamental domain $F^{\prime}$ (which is always assumed to
be closed in $\HH$), not necessarily equal to the canonical
$4g$ polygon $F$.
\item Consider a different marking
$\alpha_i^{\prime},\beta_i^{\prime}$ and a fundamental
domain $F^{\prime}$ for it.
\end{itemize}
Clearly, all the previous cases amount to an arbitary choice
of the fundamental domain for $\Gamma$. However, if $F$ and
$F^{\prime}$ are two such choices, then there exist a
suitable set of indices $\{\nu\}$, elements $\gamma_{\nu}
\in \Gamma$ and singular two-chains $c_{\nu}$ such that
\begin{equation} \label{change}
F^\prime - F =\sum_\nu(\gamma_\nu^{-1}(c_\nu) - c_\nu)\,.
\end{equation}
It follows, for instance, from the fact that the chain
complex for $\HH$ is a free
$\Gamma$-module~\cite{mac-lane}. Then we have the following

\begin{lemma} \label{invariance}
If $F$ and $F^{\prime}$ are two choices of the fundamental
domain for $\Gamma$ in $\HH$, then
$[\Sigma]=[\Sigma^{\prime}]$ for the corresponding classes
in $H_{\bullet}(\Tot\KKK_{\bullet,\bullet})$.
\end{lemma}
\begin{proof}
Let $\Sigma=F+L-V$ and
$\Sigma^{\prime}=F^{\prime}+L^{\prime}-V^{\prime}$ be the
cycles in $\Tot\KKK$ constructed according to the method
of~\ref{homology-computations}. It follows from
\eqref{change} that
\begin{equation*}
F^\prime - F = \delpp \bigl(\sum_\nu
c_\nu\otimes[\gamma_\nu]\bigr)\, ,
\end{equation*}
and therefore
\begin{equation*}
\begin{split}
F^\prime + L^\prime -F -L &= \del\bigl(\sum_\nu
c_\nu\otimes[\gamma_\nu]\bigr)\\ &\quad +\bigl(L^\prime - L
-\sum_\nu \delp (c_\nu)\otimes[\gamma_\nu]\bigr)\, .
\end{split}
\end{equation*}
The second term in these expression is an element of
$\KKK_{1,1}$ and its second differential is
\begin{equation*}
\begin{split}
\delpp \bigl(L^\prime - L -\sum_\nu \delp
(c_\nu)\otimes[\gamma_\nu]\bigr) &= \delp (F^\prime - F) -
\sum_\nu(\gamma_\nu^{-1}(\delp (c_\nu)) - \delp (c_\nu))\\
&= 0\, .
\end{split}
\end{equation*}
Since the higher homology of $\Gamma$ with values in
$\SSS_\bullet$ is zero (cf.~Appendix~\ref{app:a}), there
exists an element $C\in\KKK_{1,2}$ such that
\begin{equation*}
L^\prime - L -\sum_\nu \delp (c_\nu)\otimes[\gamma_\nu] =
\delpp C\, ,
\end{equation*}
so that
\begin{equation*}
\Sigma^\prime - \Sigma = \del \bigl(\sum_\nu
c_\nu\otimes[\gamma_\nu] -C \bigr) -V^\prime + V +\delp C\,.
\end{equation*}
Similarily, $\delpp (V^\prime - V -\delp C)=0$, and
therefore there exists $K\in\KKK_{0,3}$ such that $V^\prime
-V +\delp C=\delpp K$. Finally,
\begin{equation*}
\Sigma^\prime - \Sigma = \del \bigl(\sum_\nu
c_\nu\otimes[\gamma_\nu] - C - K\bigr)\, ,
\end{equation*}
since, obviously, $\delp K=0$.
\end{proof}
\subsubsection{Cohomology computations}
Here we pass to the dual computations in cohomology. Let
\begin{equation*}
\omega[f]=\frac{f_{zz}}{f_z}\mu_z \d z \wedge \d\bar{z}\,,
\end{equation*}
be the density of Polyakov's action functional in the genus
zero case, where $\mu=f_{\bar{z}}/f_z$. Obviously,
$\omega[f]$ can be considered as an element in $\CCC^{2,0}$,
that is a two-form valued zero cochain on $\Gamma$. Then
there exist elements $\theta[f]\in\CCC^{1,1}$ and
$\Theta[f]\in\CCC^{0,2}$ such that
\begin{equation*}
\deltapp\omega[f] =\deltap\theta[f]\quad \text{and}\quad
\deltapp\theta[f] = \deltap\Theta[f]\,,
\end{equation*}
so that the $f$-dependent cochain $\Omega_{f} \eqdef
\omega[f]-\theta[f]- \Theta[f]$ of total degree two is a
cocycle in $\Tot\CCC$, that is
\begin{equation*}
D(\omega[f]-\theta[f]-\Theta[f])=0\,.
\end{equation*}
Indeed, $\deltap\deltapp\omega[f] = \deltapp\deltap\omega[f]
=0$ because $\omega[f]$ is a top form on $\HH$, and since
\HH\ is contractible, it follows that there exists
$\theta[f]$ such that $\deltapp\omega[f]=\deltap\theta[f]$.
Similarly, $\deltap\deltapp\theta[f] =
\deltapp\deltap\theta[f] =\deltapp\deltapp\omega[f] =0$ and
again, since \HH\ is acyclic, there exists $\Theta[f]$ such
that $\deltapp\theta[f] = \deltap\Theta[f]$. Continuing
along this way, we get $\deltap\deltapp\Theta[f] = 0$, so
that $\deltapp\Theta[f]$ is a 3-cocycle on $\Gamma$ with
constant values. As it follows form
Lemma~\ref{lemma-cohomology}, $H^3(\Gamma,\CC)=\{0\}$, so
that, shifting $\Theta[f]$ by a $\CC$-valued group cochain,
if necessary, one can choose the ``integration constants''
in the equation $\d\Theta[f]=\deltapp\theta[f]$ in such a
way that $\deltapp\Theta[f]=0$.

It is quite remarkable that explicit expressions for
$\theta[f]$ and $\Theta[f]$ can be obtained by performing a
straightforward calculation. Indeed, using
\begin{equation*}
f\circ\gamma=\tilde{\gamma}\circ f\quad
\text{and}\quad\mu\circ\gamma\,
\frac{\overline{\gamma^{\prime}}}{\gamma^{\prime}}=\mu,
\end{equation*}
we get
\begin{equation} \label{omega-theta}
\deltapp\omega_{\gamma}[f]=\omega[f]
\circ\gamma^{-1}\vert(\gamma^{-1})^{\prime}
\vert^2-\omega[f]=\d \theta_{\gamma}[f].
\end{equation}
A direct computation, using the property that
$\{\gamma,z\}=0$ for all fractional linear transformations,
verifies that
\begin{equation}\label{theta}
\theta_{\gamma^{-1}}[f] = \log ({\tilde\gamma}^\prime\circ
f)\d\log f_z -\log (f_z\circ\gamma)\d\log\gamma^\prime
-2\frac{\gamma^{\prime\prime}}{\gamma^\prime}\mu\d\bar z.
\end{equation}
Proceeding along the same lines one can work out an
expression for $\Theta[f]$; in order to get a manageable
formula, it is more convenient to write down its
differential
\begin{equation}\label{Theta}
\begin{split}
\d\Theta_{\gamma_2^{-1},\gamma_1^{-1}}[f] &=
f^*\bigl(\log\bigl(\tilde\gamma_1
\circ\tilde\gamma_2\bigr)^\prime
\d\log{\tilde\gamma_2}^\prime\bigr)
+\log\gamma_2^\prime\,\d\log
\bigl(\gamma_1\circ\gamma_2\bigr)^\prime\\ &-\frac{1}{2}
f^*\bigl(\d\bigl(\log{\tilde\gamma_2}^\prime\bigr)^2\bigr)
-\frac{1}{2}\d\bigl(\log\gamma_2^\prime\bigr)^2\,.
\end{split}
\end{equation}
It is easy to verify that the right hand side of this
expression is indeed a closed one-form on $\HH$ and,
therefore, is exact.
\begin{remark}
One can obtain a formula for $\Theta[f]$ by integrating
\eqref{Theta}. The resulting expression will involve
combinations of logarithms and dilogarithms, resulting from
the typical integral
\begin{equation*}
\int\log\gamma^\prime\d\log\sigma^\prime,
\end{equation*}
where $\gamma$ and $\sigma$ are fractional linear
transformations.  The customary choice in defining this
integral is to put branch-cuts from $-\infty$ to
$\gamma^{-1}(\infty)$ and from $\sigma^{-1}(\infty)$ to
$\infty$. When these elements belong to the Fuchsian group
$\Gamma$, the branch-cuts should go along the real axis
$\RR$ which is the limit set of $\Gamma$. The same applies
to the target group $\tilde{\Gamma}$ when the mapping $f$
defines a Fuchsian deformation. If the target group
$\tilde{\Gamma}$ is quasi-Fuchsian, the branch-cuts should
go along the limit set of $\tilde\Gamma$, the simple Jordan
curve that is the image of $\RR$ under the mapping $f$. With
this normalization,
$\Theta_{\gamma_2^{-1},\gamma_1^{-1}}(f)$ is defined up to
the ``integration constants''
$c_{\gamma_2^{-1},\gamma_1^{-1}}$ which are determined from
the condition that $\deltapp\Theta[f]=0$.
\end{remark}

Therefore we proved, in complete analogy with the
homological computation, that the cochain
$\Omega_f=\omega[f] - \theta[f] -\Theta[f]\in (\Tot\CCC)^2$
is in fact a cocycle,
\begin{equation*}
D\Omega_f = 0\,.
\end{equation*}
Hence, from Lemma~\ref{lemma-cohomology}, we have
\begin{proposition}\label{cohomo-class}
The cocycle $\Omega_f\in\ (\Tot\CCC)^2$ represents a
cohomology class in $H^2(X,\CC)\cong\CC\, ,$ which depends
on the mapping $f$.
\end{proposition}
\begin{remark}
It might happen that the cohomology class $[\Omega_f]=0$ for
some specific mapping(s) $f$.
\end{remark}
\section{Polyakov's action in higher genus}
\label{action-higher-genus}
\subsection{}
After the algebraic and topological preparations of
section~\ref{algebra}, here we finally define the Polyakov
action functional and prove
Theorems~\ref{theorem:a},~\ref{theorem:b},~\ref{theorem:c}.
Let $X \simeq \Gamma\backslash\HH$ be a Riemann surface of
genus $g>1$ and $f$ be a quasi-conformal mapping such that
$\tilde{\Gamma}=f\circ\Gamma \circ f^{-1}$ is a Fuchsian or
quasi-Fuchsian group isomorphic to $\Gamma$ (see
introduction and~\ref{quasi-conformal} for details). Using
the pairing between $\CCC^{\bullet,\bullet}$ and
$\KKK_{\bullet,\bullet}$, we set
\begin{equation}\label{big-action:2}
\begin{split}
2i S[f] &= \langle\Omega_f ,\Sigma\rangle\\ & =
\langle\omega[f] ,F\rangle -\langle\theta[f],L\rangle
+\langle\Theta[f] ,V\rangle\\ & =
\int_F\omega[f]-\sum_{i=1}^g\int_{b_i}\theta_{\beta_i}[f]
+\sum_{i=1}^g\int_{a_i}\theta_{\alpha_i}[f] \\
&\quad+\sum_{i=1}^g\biggl(\Theta_{\alpha_i,\beta_i}[f](a_i(0))-
\Theta_{\beta_i,\alpha_i}[f](b_i(0))+
\Theta_{\gamma^{-1}_i,\alpha_i\beta_i}[f](b_i(0))\biggr)\\
&\qquad -\sum_{i=1}^g\Theta_{\gamma^{-1}_g \cdots
\gamma^{-1}_{i+1},\gamma^{-1}_i}[f] (b_g(0)) \, .
\end{split}
\end{equation}
\begin{proof}[Proof of Theorem~\protect\ref{theorem:a}] 
It follows at once from the constructions in
section~\ref{algebra}. First, the value of $S[f]$, for any
given $f$, depends only on the classes defined by $\Omega_f$
and $\Sigma$ and not on the explicit cocycles representing
them. Indeed, because of the property \eqref{dual} of the
pairing $\langle\;,\;\rangle$, shifting either $\Omega_f$ or
$\Sigma$ by (co)boundaries does not alter the value given
in~\eqref{big-action:2}. Furthermore, by virtue of
Lemma~\ref{invariance} and the above invariance, the action
$S[f]$ does not depend on either the choice of the marking
of $\Gamma$, or on the choice of the fundamental domain $F$.
Finally, it follows from Propositions~\ref{fund-class}
and~\ref{cohomo-class}, which identify the (total) homology
of the complexes $\KKK_{\bullet,\bullet}$ and
$\CCC^{\bullet,\bullet}$ with that of the surface $X$, that
the action $S[f]$ comes from the pairing
\begin{equation*}
H^2(X,\CC)\times H_2(X,\ZZ)\longrightarrow \CC\,.
\end{equation*}
\end{proof}
\begin{remark}
Since the action results from a pairing in homology, we
write it as
\begin{equation}
S[f] = \frac{1}{2i}\langle [\Omega_f],[\Sigma]\rangle,
\end{equation}
stressing its dependence on the (co)homology classes only.
\end{remark}
\subsection{} \label{subsection-variation}
Here we discuss the variational properties of the action
functional~\eqref{big-action:2} and prove
Theorem~\ref{theorem:b}.  As it was mentioned in the
introduction, there are two versions of the variational
problem for $S[f]$. In the first one, the free-end
variation, we consider $\mu$ to be the independent variable,
so that the target Fuchsian (or quasi-Fuchsian) group
$\tilde{\Gamma}$ is determined by $\mu$ through the solution
of the Beltrami equation. In the second case, the fixed-end
variation, we fix the target Fuchsian (or quasi-Fuchsian)
group $\tilde{\Gamma}$, together with the isomorphism
$\Gamma \longrightarrow \tilde{\Gamma}$ and consider the set
$\qc$ of all smooth quasi-conformal mappings $f$ that
intertwine between $\Gamma$ and $\tilde{\Gamma}$.

In the first case, since the set of Beltrami coefficients
for $\Gamma$ is the interior of a ball of radius $1$ (with
respect to the $\vert\vert~\vert\vert_{\infty}$ norm) in the
linear space $\mathcal{B}(\Gamma)$ of all Beltrami
differentials for $\Gamma$, the variation $\var\mu$ belongs
to $\mathcal{B}(\Gamma)$.

In the second case, since the target Fuchsian (or
quasi-Fuchsian) group $\Gamma$ is fixed, it follows from the
equivariance property \eqref{eq} that $\var f/f_z$ is
$(-1,0)$-tensor for $\Gamma$, that is
\begin{equation*}
\frac{\var f}{f_z}\circ\gamma = \frac{\var
f}{f_z}\,\gamma^\prime \quad\text{for all
$\gamma\in\Gamma$.}
\end{equation*}
One can express $\var f/f_z$ in terms of a vector field on
$X$ as follows.  Let $\mathcal{G}_0$ be the group of all
orientation preserving diffeomorphisms of \HH\ fixing
$\Gamma$ and homotopic to the identity. Any path $g^t$ in
$\mathcal{G}_0$ connected to the identity defines a path
$f^t = f\circ g^t$ in $\qc$ connected to $f\in\qc$, a
deformation of the mapping $f$. Setting
\begin{equation*}
\var f = \frac{d}{dt}\biggr|_{t=0}\, f^t
\end{equation*}
and defining $v=v^z\,\del_z+v^{\bar z}\,\del_{\bar z}$ as
the vector field generating the flow $t\mapsto g^t$, we get
\begin{equation*}
\frac{\var f}{f_z} = v^z +\mu\,v^{\bar z}\, ,
\end{equation*}
where $\mu=f_{\bar{z}}/f_z$ is the Beltrami coefficient for
$\Gamma$ corresponding to $f$.

Note that in the first case the corresponding variation
$\var f/f_z$ is not necessarily a $(-1,0)$-tensor for
$\Gamma$, since the target group $\tilde{\Gamma}$ ``floats''
under a generic variation of $\mu$ (variation with free
end). Specifically, 
\begin{equation} \label{eichler}
\frac{\var f}{f_z}\circ \gamma\, \frac{1}{\gamma^{\prime}}=
\frac{\var f}{f_z} +
\frac{1}{f_z} \biggl(
\frac{\var\tilde\gamma}{\tilde\gamma^{\prime}} \biggr) 
\circ f\,,
\end{equation} 
for all $\gamma\in\Gamma$. Objects on $\HH$ with such
tranformation property are pull-backs under the map $f$ of
non-holomorphic Eichler integrals of order $-1$ for the
group $\tilde\Gamma$. By definition~\cite{kra}, the space
$\EEE^{-1}_{\tilde\Gamma}$ of these Eichler integrals
consists of smooth functions $\mathcal{E}$ on $\HH$ such
that
\begin{equation} \label{Eichler}
\mathcal{E}\circ\tilde\gamma\,
\frac{1}{\tilde\gamma^{\prime}} =
\mathcal{E}+p_{\tilde\gamma}\,,
\end{equation}
for all $\tilde\gamma\in\tilde\Gamma$, where
$p_{\tilde\gamma}$ is a $1$-cocycle of $\tilde\Gamma$ with
coefficients in the linear space of polynomials $P$ of order
$\leq 2$ with the action $$P\mapsto
((\tilde\gamma^{-1})^{\prime})^2 P\circ\tilde\gamma^{-1}.$$
Clearly the pull-back $(\mathcal{E}\circ f)/f_z$ of the
Eichler integral $\mathcal{E}$ has the trasformation
property~\eqref{eichler}.
 
In both cases the variations of $f$ and $\mu$ are related by
the same equation
\begin{equation*}
\diff{M}\biggl(\frac{\var f}{f_z}\biggr)=\var\mu,
\end{equation*}
where $\diff{M}=\delb -\mu\,\del +\mu_z$ is the differential
operator introduced in section~\ref{genus-zero}. It has the
remarkable property of mapping $(-1,0)$-tensors for $\Gamma$,
and even objects of more complicated type such as pull-backs of 
Eichler integrals, into $(-1,1)$-tensors for $\Gamma$. There
are other differentials operators with similar properties,
collected in the following 
\begin{lemma}\label{tensors}
\begin{itemize}
\item[(i)] The operators $\diff{T}=\del^3+2T\del+T_z$ and
$\diff{M}=\delb-\mu\del+\mu_z$, where $T$ is a quadratic
differential for $\Gamma$ and $\mu$ is Beltrami differential
for $\Gamma$, map $(-1,0)$-tensors for $\Gamma$ into
quadratic and Beltrami differentials for $\Gamma$,
respectively.
\item[(ii)] The operators $\mathcal{T}$ and $\mathcal{M}$ from part
\textrm{(i)} map pull-backs by the mapping $f$ of Eichler
integrals of order $-1$ for $\tilde\Gamma$ into quadratic
and Beltrami differentials for $\Gamma$, respectively.
\item[(iii)] If $f$ is mapping of $\HH$ intertwining $\Gamma$
and $\tilde{\Gamma}$, then $T=\{f,z\}$ is a quadratic
differential for $\Gamma$.
\end{itemize}
\end{lemma} 
\begin{proof} Part (i) is well-known (see, e.g.~\cite{gun}) and
the statements can be easily verified. In particular,
setting $T=0$ we get that $\mu_{zzz}$ is a $(2,1)$-tensor
for $\Gamma$, which is also a known result (see,
e.g.~\cite{kra}). 

In order to prove part (ii), note that for
a holomorphic function $p$ on $\HH$ we have
\begin{equation*}
\mathcal{T}\biggl(\frac{p\circ f}{f_z}\biggr)= f^{2}_{z}\,
(\del^3 p)\circ f\,,
\end{equation*}
which shows that the additional terms in the transformation
law \eqref{eichler} belong to the kernel of
$\mathcal{T}$. Similarly, \eqref{delbar} shows that these
terms belong to the kernel $\mathcal{M}$ as well.

Part (iii) is another classical result,
which can be easily verified as well.
\end{proof} 
\subsubsection{Proof of Theorem~\protect\ref{theorem:b}}
For concreteness, we first consider variations with respect
to $\mu$, though, as we shall see, the actual argument works
for both kinds of variations.

The proof requires climbing the ``ladder'' in the double
complex $\CCC^{\bullet,\bullet}$, together with the
computation of the variation of $\omega[f]$. Since
$\omega[f]$ is a local functional of $f$, we can just use
the computation already done in genus zero so that,
according to formula \eqref{varomega},
\begin{equation}\label{variation}
\var\omega = a - \d\eta\, ,
\end{equation}
where $a=-2\,T\,\var\mu\d z\wedge\d\bar{z}$ and the explicit
expression for the $1$-form $\eta$ is not needed. (In order
to simplify notations, we temporarily drop the dependence on
$f$ from the notation.) As it follows from
Lemma~\ref{tensors}, the $2$-form $a$ on $\HH$ is a
$(1,1)$-tensor for $\Gamma$, therefore it is closed with
respect to the total differential, i.e.~$Da=0$.

Next observe that $D\var\Omega =\var D\Omega =0$, therefore
$D(\var\Omega - a)=0$. We want to show that $\var\Omega - a$ is in
fact $D$-exact up to a term whose contribution vanishes
after pairing with $\Sigma$.

To this end, let us write
\begin{equation*}
\var\Theta = \deltapp\chi\,,
\end{equation*}
where $\chi$ has degree $(0,1)$ in the total complex. This
is possible, since, as it is shown in the appendix, the
higher cohomology of $\Gamma$ with coefficients in the de
Rham complex vanishes. The equation $D\var\Omega =0$ gives
us the two relations
\begin{equation} \label{*}
\deltap\var\Theta =\deltapp\var\theta\,,\quad
\deltap\var\theta = \deltapp\var\omega\,,
\end{equation}
of which the first one implies that
\begin{equation*}
\var\theta = \deltap\chi + \deltapp\lambda\,,
\end{equation*}
where, again, the vanishing of $H^q(\Gamma,\AAA^p)$ for
$q>0$ has been used. Plugging this relation into the second
one in \eqref{*}, yields
\begin{equation*}
\deltapp\var\omega=\deltapp\deltap\lambda\,.
\end{equation*}
Notice that this time we can at most conclude that
$\var\omega-\deltap\lambda$ is a $\Gamma$-invariant form,
since $H^0(\Gamma,\AAA^p)$ precisely gives the invariant
$p$-forms (cf.~the appendix). We write this invariant form
as $a+b$, for some $(2,0)$ invariant element $b$, so that
\begin{gather*}
\var\omega = \deltap\lambda + a + b\\ \intertext{and, using
\eqref{variation},} b=-\deltap (\eta +\lambda)\,,
\end{gather*}
i.e.~ $b$ is $\Gamma$-invariant and exact. Putting all
together, we obtain
\begin{align*}
\var\Omega &= \var\omega - \var\theta - \var\Theta\\ &= a -
\deltap\eta - \deltap\chi - \deltapp\lambda - \deltapp\chi\\
&= a + b + D (\lambda - \chi)\,,
\end{align*}
which, after evaluation against $\Sigma$, reduces to
\begin{equation*}
\langle\var\Omega\,,\,\Sigma\rangle = \int_F a\,,
\end{equation*}
as wanted (the integral of $b$ over $F$ is obviously zero).

In order to complete the proof, notice that the variation of
$\omega[f]$ always has the form \eqref{variation},
independently of whether either variable $\mu$ or $f$ is
varied. In the latter case, the variation $\var f/f_z$ is a
$(-1,0)$-tensor for $\Gamma$, so that we can use
\eqref{variation} and the relation $\var\mu = \diff{M}(\var
f/f_z)$ together with Lemma~\ref{anom}.\qed

\begin{remark} Note that the argument presented in the
proof of Theorem~\ref{theorem:b} is quite general. It
applies to any functional defined by an evaluation of a
cocycle in $\Tot\CCC^2$ over a cycle $\Sigma$, provided that
the cocycle is the extension of a $2$-form on $\HH$ with the
property that its variation is a sum of $D$ and $\d$-exact
terms.
\end{remark} 
\subsubsection{}
\label{two-complex-structures}
As it was mentioned in the introduction, it follows from
Theorem~\ref{theorem:b} that $c\,S[f]/24\pi$, considered as
a functional of $\mu=f_{\bar{z}}/f_z$, solves equation
\eqref{WI}, no matter what kind of deformation we are
considering, be it Fuchsian or quasi-Fuchsian. Thus there
are at least two possible solutions of \eqref{WI} on a
Riemann surface of genus higher than one. In order to
clearly distinguish the two cases, let us adopt for a moment
the customary notation in the theory of quasi-conformal
mappings~\cite{ahl}, so that $f^\mu$ and $\Gamma^\mu$
(respectively $f_\mu$ and $\Gamma_\mu$) stand for the
Fuchsian (respectively, quasi-Fuchsian) deformation of
$\Gamma$.

There is a simple relationship between the variations of
$S[f_\mu]$ and $S[f^\mu]$. First of all, observe that the
mapping $g:=f_\mu\circ(f^\mu)^{-1} : \HH \rightarrow
f_\mu(\HH)$ is conformal (note that $f^\mu(\HH)=\HH$).
Indeed, it follows from the Beltrami equation that
\begin{equation*}
\frac{\del g}{\del\bar{\zeta}} = \frac{\del f^\mu}{\del z}
\biggl(\frac{\del (f^\mu)^{-1}}{\del\bar{\zeta}} + \mu\,
\frac{\del\overline{(f^\mu)^{-1}}}{\del\bar{\zeta}}
\biggr)=0\,,
\end{equation*}
where $\zeta=f^\mu(z,\bar{z})$ is the new complex coordinate
on $\HH$. Moreover, the map $g$ intertwines $\Gamma^\mu$ and
$\Gamma_\mu$, thus it descends to a biholomorphic map
\begin{equation*}
g: X^{\mu}=\Gamma^\mu\backslash\HH \longrightarrow
\Gamma_\mu\backslash f_\mu(\HH)=X_{\mu}
\end{equation*}
showing that the Riemann surfaces $X^{\mu}$ and $X_{\mu}$
are conformally equivalent.  Furthermore, we have
\begin{equation*}
T_\mu(z)=\{f_\mu,z\}=\{g,\zeta\}\circ f^\mu\,(f^\mu_z)^2 +
T^\mu(z)\,,
\end{equation*}
where $T^\mu(z)=\{f^\mu,z\}$. Thus the difference
\begin{equation*}
Q=\frac{\var S[f_\mu]}{\var\mu} - \frac{\var
S[f^\mu]}{\var\mu}
\end{equation*}
is just the pull-back under $f^\mu$ of the holomorphic
quadratic differential obtained by taking the Schwarzian
derivative of $g$ with respect to the new complex coordinate
$\zeta$. Of course, the situation is completely symmetric
under the exchange of $f_\mu$ and $f^\mu$.

One can reach the same conclusion proceeding along a
different line (cf.~\cite{yoshida}). Namely, since both
$S[f^\mu]$ and $S[f_\mu]$ satisfy \eqref{WI}, $Q$ satisfies
the equation
\begin{equation*}
(\delb -\mu\,\del -2\,\mu_z)Q=0
\end{equation*}
which, using the Cauchy-Riemann operator
\begin{equation*}
\frac{\del}{\del\bar\zeta} =\frac{\del\bar
z}{\del\bar\zeta}\,\biggl(\frac{\del}{\del\bar z}-
\mu\,\frac{\del}{\del z}\biggr)
\end{equation*}
can be written as
\begin{equation*}
\del_{\bar\zeta}\biggl( \frac{Q}{f_z^2}\biggr) = 0\,,
\end{equation*}
showing that $Q$ is indeed the pull-back of a holomorphic
quadratic differential with respect to the complex
coordinate $\zeta$.
\begin{remark} The above argument actually shows that 
homogeneous solutions to the equation~\eqref{WI} on $X$ are
pull-backs under the mapping $f^{\mu}$ (or $f_{\mu}$) of the
holomorphic quadratic differentials on the ``target''
Riemann surface $X^{\mu}$. According to the Riemann-Roch
theorem, this space is $3g-3$-dimensional; therefore, the
universal CWI~\eqref{WI} does not completely determine the
generating functional for the stress-energy tensor in the
higher genus case.  As we mentioned in the introduction,
additional information should be provided by the particular
CFT. 
\end{remark}
\subsubsection{} 
According to Theorem~\ref{theorem:b}, the variation of the
action with respect to the map $f$ yields the classical
equation of motion
\begin{equation}\label{critical}
\mu_{zzz}=0\, .
\end{equation}
Here we compute the dimension of the space of solutions
of~\eqref{critical}. It was observed in the introduction
that determining the critical set of $S[f]$ in $\qc$ out
of~\eqref{critical} seems to be a very difficult problem.
However, the space of solutions to \eqref{critical} is quite
interesting since, as we show below, it contains the
subspace of harmonic Beltrami differentials.

First, recall the definition of the so-called Maass
operators (see, e.g.~\cite{wolpert}). For $k,l\in\ZZ$,
denote by $\AAA^{k,l}_\Gamma\equiv A^{k,l}_\CC(\HH)^\Gamma
\cong A^{k,l}_\CC(X)$ the space of $\Gamma$-invariant
$(k,l)$-forms on $\HH$; by convention, $(\d\! z)^k$, for $k$
negative, means $(\del /\del z)^{-k}$. Define
\begin{gather*}
D_{k,l}:\AAA^{k,l}\longrightarrow \AAA^{k+1,l}\\
\intertext{by} D_{k,l}= y^{-2k}\circ\del\circ y^{2k}\, ,
\end{gather*}
where $\del=\del/\del z$.  It is easy to verify that
\begin{equation}\label{del-cube}
\del_z^{\, 3} = D_{1,1}\circ D_{0,1}\circ D_{-1,1}\, ,
\end{equation}
which once again shows that the operator $\del_z^{\, 3}$
maps Beltrami differentials into the $(2,1)$-tensors for
$\Gamma$. Furthermore, a Beltrami differential
$\nu\in\AAA^{-1,1}_\Gamma$ is called Bers harmonic if it is
harmonic with respect to the $\del$-Laplacian of the
Poincar\'{e} metric on $\Gamma\backslash\HH$, acting on
$(-1,1)$-forms. It can be shown that
\begin{equation*}
\nu = y^2\,\bar q\,,
\end{equation*}
where $q\in\AAA^{2,0}_\Gamma$ is a holomorphic quadratic
differential. It follows from the Riemann-Roch theorem that
Bers harmonic Beltrami differentials form a
$(3g-3)$-dimensional complex vector space and play an
important role in the Teichm\"{u}ller theory~\cite{ahl,gar}.
\begin{proposition}\label{proposition:critical}
The space of solutions of equation~\eqref{critical} has
complex dimension $4g-3$:
\begin{equation*}
\dim_\CC\Ker_{\AAA^{-1,1}_\Gamma}(\del_z^{\, 3}) = 4g -3,
\end{equation*}
and contains the $3g-3$ dimensional vector space of Bers
harmonic Beltrami differentials.
\end{proposition}
\begin{proof} 
Using \eqref{del-cube}, we start by observing that the
kernel of $D_{-1,1}$ coincides with the space of harmonic
Beltrami differentials. Indeed, $\nu\in\Ker (D_{-1,1})$ if
and only if $\del (y^{-2}\nu )=0$, which implies $\nu =
y^2\,\bar q$, for $q$ a holomorphic quadratic differential,
since $y^{-2}\nu$ is a $(0,2)$-form.

Furthermore, $\Ker (D_{1,1})\cap\im (D_{0,1})
=\{0\}$. Indeed, an element in $\Ker (D_{1,1})$ is
necessarily a multiple of the $(1,1)$-form $y^{-2}$. If it
is non zero, then it cannot belong to $\im
(D_{0,1})=\im\del$, since $y^{-2}$ represents a cohomology
class in $\Gamma\backslash\HH$.

Next, it is clear that $\Ker (D_{0,1})$ is complex
anti-isomorphic to the linear space of Abelian differentials
for $X$. Finally, the map $D_{-1,1}$ is onto: its image is
the entire space of $(0,1)$-differentials. Namely, the
operator adjoint to $D_{-1,1}$ with respect to the Hermitian
scalar product on $\AAA^{k,l}_\Gamma$ induced by the
Poincar\'{e} metric $y^{-2}$ is $D_{-1,1}^* = -\delb\circ
y^2$, which has zero kernel since $g>1$. Thus any element in
$\Ker (D_{0,1})$ is the $D_{-1,1}$-image of an element in
$\AAA^{-1,1}_\Gamma$, orthogonal to the subspace of harmonic
Beltrami differentials, and it also belongs to the kernel of
$\del_z^{\,3}$. Counting $4g-3=3g-3+g$ proves the claim.
\end{proof}

\begin{remark} 
As in the genus zero case, the equation of motion
\eqref{critical} is equivalent to the holomorphicity
property of $T=\{f,z\}$ with respect to the new complex
structure induced by $f$.
Namely, when $\mu$ satisfies~\eqref{critical}, the
corresponding \eqref{WI} becomes homogeneous so that,
according to~\ref{two-complex-structures}, we have
\begin{equation}\label{critical-holo}
\del_{\bar\zeta}\biggl( \frac{T}{(\del_z\zeta)^2}\biggr) = 0
\end{equation}
for the stress-energy tensor in the new coordinates $\zeta\,
,\bar\zeta$. This condition is well defined on the surface
$X$ as well as on the deformed Riemann surface
$\tilde\Gamma\backslash f(\HH)$.
\end{remark}
\subsubsection{}
Here we briefly comment on the computation of the second
variation. It follows from Lemma~\ref{tensors} that the
differential operators used in the genus zero computation
are tensorial; therefore, using Theorem~\ref{theorem:b} and
the fact that the problem is local, we can just repeat the
computations in~\ref{second-variation} in order to get the
\begin{proposition}
The Hessian of the Polyakov action \eqref{big-action:2} is
given by the genus zero formula
\begin{equation*}
\var^2\, S[f](\var_1f,\var_2f) =-2\int_F
\frac{\var_1f}{f_z}\bigl(\del^3\circ\diff{M}\bigr)
\biggl(\frac{\var_2f}{f_z}\biggr)\, \d^2z\, .
\end{equation*}
\end{proposition}
\subsection{}
We now analyze how $S[f]$ relates to the functional $W[\mu]$
defined by \eqref{WW}, and prove Theorem~\ref{theorem:c}.

For $t\in [0,1]$, let $\mu^t$ be a homotopy in the space of
Beltrami differentials connecting $0$ to $\mu$, and let
$f^t$ the solution of the Beltrami equation corresponding to
$\mu^t$.  For the sake of convenience, let us rewrite
\eqref{WW} here:
\begin{equation}\label{WWW}
W[\mu]=\frac{c}{12 \pi} \int^1_0\biggl(\int_{F} T^t
\,\dot{\mu}(t)\,\d^2 z\biggr)\d t\,.
\end{equation}
The integration in \eqref{WWW} is extended to $F$, but,
according to Lemma~\ref{tensors}, the integrand is a
$(1,1)$-tensor for $\Gamma$, hence the integral descends to
$X$.

\begin{proof}[Proof of Theorem~\protect\ref{theorem:c}]
We want to proceed in a fashion similar to the proof of
Theorem~\ref{theorem:b}.

Our construction of $S[f]$ applied to $f^t$ produces
$\omega^t$, $\Omega^t$ and $S[f^t]$ for any $t\in [0,1]$. We
can make use of formula \eqref{varomega} applied to $\var
=d/dt$:
\begin{equation*}
\dot\omega^t = -2\, T^t\dot\mu^t\,\d z\wedge\d\bar z -
\d\eta(f^t;\dot f^t)\equiv a^t - \d\eta^t\, ,
\end{equation*}
where, as before, $D a^t=0$. On the other hand, $D
\dot\Omega^t=0$, since $D\Omega^t=0$ for any $t$, and
therefore the same arguments as in the proof of
Theorem~\ref{theorem:b} lead us to conclude that
\begin{equation*}
\langle \dot\Omega^t,\Sigma\rangle =\int_F a^t\, .
\end{equation*}
Integrating in $t$ from $0$ to $1$ we get that
$W[\mu]=(c/24\pi ) S[f]$, which together with Theorem~\protect
\ref{theorem:c} proves part (i)..

First statement of the part (ii) follows from the is well-known~\cite{ahl} 
that the quasi-Fuchsian deformation $f=f_{\mu}$ depends holomorphically on 
$\mu$. Finally, if $f=f^{\mu}$ is
a Fuchsian deformation with harmonic Beltrami differential
$\mu=y^2\bar{q}$, then the Ahlfors lemma (see,
e.g.,~\cite{zt}) states
\begin{equation*}
\frac{\del f^{\epsilon\mu}}{\del\bar\epsilon}
\bigg|_{\epsilon=0}=-\frac{1}{2}q.
\end{equation*}
Therefore, choosing a linear homotopy $\mu(t)=t\mu$, we have the following  
simple computation 
\begin{align*}
\frac{\del^2 W[\epsilon\mu]}{\del\epsilon\del\bar\epsilon}
\bigg|_{\epsilon=0}&=\frac{c}{12\pi}\int_{0}^{1}\int_F\frac{\del f^{\epsilon 
t\mu}}{\del\bar\epsilon}\bigg|_{\epsilon=0}\mu \d^2 z \d t \\
& -\frac{c}{24\pi} \int_{0}^{1}t\d t \int_F q\mu\d^2 z \\ 
&= -\frac{c}{48 \pi}\int_F |\mu|^2
y^{-2}\d^2z.
\end{align*}
\end{proof}
\begin{remark}
Theorem~\ref{theorem:c} specifies the $\mu$-dependence for
two natural solutions for $W[\mu]$, defined by quasi-Fuchsian
and Fuchsian deformations. In the former case the
corresponding functional is holomorphic in $\mu$, as a
generating functional should be, while in the latter case it
is not. Introducing the Weil-Petersson inner product in the
space of Bers harmonic Beltrami differentials by
\begin{equation*}
\big(\mu_1,\mu_2\big)_\mathrm{WP}=
\int_F\mu_1{\bar\mu}_2\,y^{-2}\d^2z,
\end{equation*}
the latter statement takes a quantative form
\begin{equation*}
\frac{\del^2 W[\epsilon\mu]}{\del\epsilon\del\bar\epsilon}\bigg|_{\epsilon=0}=
-\frac{c}{48\pi}||\mu||^{2}_\mathrm{WP},
\end{equation*}
that once again characterizes the Weil-Petersson metric as a
``holomorphic anomaly''. Finally, for arbitary Beltrami
differential one should replace $\mu$ by $P\mu$ in the above
formula, where $P$ stands for the orthogonal projection
(with respect to the Weil-Petersson metric) onto the space
of harmonic Beltrami differentials.
\end{remark}
\subsection{}
Here we compute the Hessian of the action functional $W$ as 
a functional of $\mu$. For this end we need to
extend the linear mapping $\mathcal{M}: \AAA^{-1,0}_{\Gamma}
\rightarrow \AAA^{-1,1}_{\Gamma}$ to the space of pull-backs
by the mapping $f$ of Eichler integrals of order $-1$ for 
$\tilde\Gamma$. This mapping has no kernel on the subspace of
normalized Eichler integrals (i.e.~vanishing at $0,1,\infty$)
and, according to Bers, it is onto (see~\cite{kra}). We denote,
slightly abusing the notations, the inverse of thus extended
mapping $\mathcal{M}$ by $\mathcal{M}^{-1}$.
 
\begin{proposition}
The second variation of the functional $W[\mu]$ is given by
\begin{equation*}
\var^2\, W[\mu](\var_1\mu,\var_2\mu)=\frac{c}{12\pi}\int_F
\var_1\mu\, \bigl(\mathcal{T}\circ\mathcal{M}^{-1}\bigr)\, 
(\var_2\mu)\, \d^2z\,,
\end{equation*}
where, according to Lemma~\ref{tensors}, the operator
$\mathcal{T}\circ\mathcal{M}^{-1}$ maps Beltrami
differentials for $\Gamma$ into quadratic
differentials. The Hessian of $W[\mu]$ at the point $\mu$ is given by the
operator $\del^3\circ\mathcal{M}^{-1}$.
\end{proposition}
\begin{proof}
It is the same as the genus zero computations using
Lemma~\ref{tensors}. Note that at the critical point
$T(z)=0$, so that $\mathcal{T}=\del^3$.
\end{proof} 

\section{Fiber spaces over Teichm\"uller space. Discussion
and conclusions}
\label{fiber-spaces}
In the preceding sections we have defined the Polyakov's
action for the chiral sector in the induced gravity on a
Riemann surface $X$ of genus $g>1$ and explored some of its
properties. We have also pointed out the possible
interpretation of $W[\mu]=(c/24\pi)\,S[f]$ as the universal
part of the generating functional for the correlation
functions of the stress-energy tensor for a CFT on $X$.

However, the most compelling interest in $W[\mu]$ (or
$S[f]$) stems in its relation with the geometry of the
various fiber spaces over Teichm\"uller space. We want to
elaborate more on this point. 
\subsection{}
Recall that the Teichm\"uller space $\mathcal{T}(X)$ of the
Riemann surface $X$ of genus $g>1$ is naturally realized as
the quotient of the open unit ball $\mathcal{B}(X)$ (with
respect to the $L^{\infty}$ norm) in the Banach space of
Beltrami differentials on $X=\Gamma\backslash\HH$ by the
group of quasi-conformal self-mappings of $\HH$ pointwise
fixing the group $\Gamma$. If one replaces $\mathcal{B}(X)$
by its subset $\mathcal{P}(X)$ consisting of smooth Beltrami
differentials and considers the identity component
$\mathcal{G}_{0}(X)$ of the group $\mathcal{G}(X)$ of
orientation preserving diffeomorphisms of $X$ (elements in
$\mathcal{G}_0(X)$ point-wise fix $\Gamma$ while acting on
$\HH$), then one gets Earle and Eells~\cite{ee} fiber space
$\pi :\mathcal{P}(X) \rightarrow \mathcal{T}(X)$ over the
Teichm\"{u}ller space.  It is a smooth (in the Frech\'{e}t
topology) principal $\mathcal{G}_0(X)$-bundle over
$\mathcal{T}(X)$. The group action on $\mathcal{P}(X)$ can
be written as $\mu = \mu(f)\mapsto \mu^g = \mu(f\circ g)$,
for $g\in \mathcal{G}_0(X)$~\cite{ee}, where $f=f^{\mu}$ is a
Fuchsian deformation associated with $\mu$. Explicitly, the 
above action is~\cite{ahl}:
\begin{equation*}
\mu^g = \frac{\overline{g_z}}{g_z} \biggl(\frac{\mu - \mu
(g^{-1})}{1 - \mu\,\overline{\mu (g^{-1})}}\biggr) \circ
g\,.
\end{equation*}

Consider now the tangent bundle exact sequence
\begin{equation*}
0\longrightarrow T_V\mathcal{P}(X) 
\overset{i}{\longrightarrow}
T\mathcal{P}(X) 
\overset{\d\pi}{\longrightarrow} \pi^*(T\mathcal{T}(X))
\longrightarrow 0
\end{equation*}
determined by the Earle-Eells fibration. (Observe that since
$\mathcal{P}(X)$ is a ball in the vector space
$\AAA^{-1,1}_\Gamma$ of all smooth Beltrami differentials,
the tangent space to it at any given point $\mu$ is
canonically identified with $\AAA^{-1,1}_\Gamma$.) According
to the description of the fixed-end variation given
in~\ref{subsection-variation}, the deformation $f^t=f\circ
g^t$, for $t\mapsto g^t\in\mathcal{G}_0(X)$, results in a
vertical curve $t\mapsto \mu^t$ above the point $\pi
(\mu)\in \mathcal{T}(X)$. Thus the corresponding variation
$\var\mu=\dot{\mu}$ lies in the vertical tangent space
$T_V\mathcal{P}(X)$ at point $\mu$, which is isomorphic to
$\im(\mathcal{M})$, where $\mathcal{M}=\delb-\mu\del+\mu_z
:\AAA^{-1,0}_\Gamma \rightarrow \AAA^{-1,1}_\Gamma$.  Next,
the tangent space $T_{\mu}\mathcal{P}(X)$ can also be
identified with the space of smooth $\tilde\Gamma$-Beltrami
differentials; an easy computation proves the following
(well-known) lemma.
\begin{lemma}
For any $\nu\in\AAA^{-1,1}_\Gamma$ the correspondence
\begin{equation*}
\nu \mapsto \biggl(\frac{f_z}{\overline{f_z}} \frac{\nu}{1 -
\vert\mu\vert^2}\biggr)\circ f^{-1}
\end{equation*}
maps $\AAA^{-1,1}_\Gamma$ isomorphically onto
$\AAA^{-1,1}_{\tilde\Gamma}$. Under this map $\mathcal{M}$
becomes $\delb_{\pi (\mu)}$, the $\delb$-operator relative
to the new complex structure on the Riemann surface $X$
defined by $\mu$.
\end{lemma}
This implies at once that the kernel of $\mathcal{M}$ is
trivial, and therefore the correspondence
\begin{equation*}
v=v^z\del_z +v^{\bar{z}}\del_{\bar{z}}\mapsto
\mathcal{M}(v^z +\mu v^{\bar{z}})
\end{equation*}
explicitly gives the injection in the tangent bundle
sequence above.  Furthermore, it realizes
$T_V{\mathcal{P}(X)}$ (and its quotient by
$\mathcal{G}_0(X)$) as a bundle of Lie algebras, as usual in
a principal fibration~\cite{atiyah}. Here the Lie algebra in
question is the Lie algebra $\mathrm{Vect}(X)$ of smooth
vector fields on $X$, which can be identified---as a real
vector space---with $\AAA^{-1,0}_\Gamma$.

With these definitions at hand, the following
reinterpretation of the formulas in the statement of
Theorem~\ref{theorem:b} becomes obvious.
\begin{proposition}\label{restatement}
For any smooth functional $\mathcal{F}:\mathcal{P}(X)\rightarrow \CC$,
\begin{enumerate}
\item the open-end variation $\var\mathcal{F} $ computes its total
differential on $\mathcal{P}(X)$;
\item the fixed-end variation computes its vertical
differential.
\end{enumerate}
In particular, for the action functional $W$,
\begin{equation*}
\d W|_{\mu}=\frac{c}{12\pi}T\in T^{*}_{\mu}\mathcal{P}(X).
\end{equation*}
\end{proposition}
\begin{remark}
The second point in the proposition can be verified by the  
following explicit computation, that uses 
Theorems~\ref{theorem:b},~\ref{theorem:c} and Lemma~\ref{anom}.
\begin{align*}
\frac{\var W}{\var f(z)} &=
-\frac{c}{12\pi}\int_F\mu_{zzz}\, \frac{\var f}{f_z}\,\d^2z
=-\frac{c}{12\pi}\int_F\mathcal{D}T(z)\, \frac{\var
f}{f_z}\, \d^2z \\
&= \frac{c}{12\pi}\int_F T(z)\mathcal{M} \biggl(\frac{\var
f}{f_z}\biggr)\,\d^2z.
\end{align*}
\end{remark}
\begin{remark}
The description of the vertical bundle as the image of
$\mathcal{M}$ immediately implies that
\begin{equation*}
T_{\pi(\mu)}\mathcal{T}(X)\cong
\AAA^{-1,1}_\Gamma/\im(\mathcal{M})\,, 
\end{equation*}
so that we get the well-know result~\cite{ee}
\begin{equation*}
T_{\pi(\mu)}\mathcal{T}(X)\cong H^{0,1}_{\delb}
(X^\mu,T_{X^{\mu}}) \cong H^1(X^\mu , \Theta_{X^{\mu}})\, ,
\end{equation*}
where the last group gives the Kodaira-Spencer infinitesimal
deformations. ($\Theta_{X^{\mu}}$ is the holomorphic tangent
sheaf to the Riemann surface $X^{\mu}$.)
\end{remark}

\subsection{}
It is fundamental to investigate how the function
$W:\mathcal{P}(X) \rightarrow \CC$ relates to the geometry
of the bundle $\pi :\mathcal{P}(X)\rightarrow
\mathcal{T}(X)$. A long but straightforward computation
using the definition \eqref{WW} of $W$ proves
\begin{lemma}
There exists $A:\mathcal{P}(X)\times \mathcal{G}_0(X)
\rightarrow \CC$ such that
\begin{equation}\label{pol-wieg:1}
W[\mu^g]  = W[\mu] + A[\mu,g]\,.
\end{equation}
The functional $A$ depends only on the point $(\mu ,g)$ and
is local in $\mu$ and $\mu^g$; in particular, it is
independent of any possible choice of the solution of the
Beltrami equation involved in the definition of $W$.
\end{lemma}
It trivially follows from \eqref{pol-wieg:1} that the
functional $A$ satisfies the cocycle identity:
\begin{equation*}
A[\mu,gh]=A[\mu^g,h]+A[\mu,g]\,.
\end{equation*}

Next, according to~\cite{ver}, the functional
$\Psi[\mu]=\exp(-W[\mu])$ is to be interpreted as a
conformal block for a CFT defined on $X$. Thus it is more
convenient to work with the exponential version of
\eqref{pol-wieg:1}. Namely, defining 
\begin{equation*}
C[\mu,g]=\exp (-A[\mu,g])\,,
\end{equation*}
we get 
\begin{equation}\label{pol-wieg:2}
\Psi[\mu^g] = C[\mu,g]\,\Psi[\mu]\,.
\end{equation}
The cocycle condition takes the form 
\begin{equation*}
C[\mu,gh]=C[\mu^g,h]\,C[\mu,g]\,,
\end{equation*}
which defines a 1-cocycle on $\mathcal{G}_0(X)$ with values
in the group of non vanishing complex valued functions on
$\mathcal{P}(X)$. We denote by $[C]$ the class of $C$ in the
cohomology group $H^1(\mathcal{G}_0(X),
\CC^*(\mathcal{P}(X)))$. 
\begin{proposition}
There is an injective map of the group
$H^1(\mathcal{G}_0(X), \CC^*(\mathcal{P}(X)))$ into the
group of isomorphism classes of line bundles over
$\mathcal{T}(X)$. The line bundle $L_{[C]}$ over $\mathcal{T}(X)$, 
defined by $[C]$ is, in particular, holomorphic.
\end{proposition} 
\begin{proof}
The existence of a map
\begin{equation*}
0\rightarrow H^1(\mathcal{G}_0(X), \CC^*(\mathcal{P}(X)))
\rightarrow H^2(\mathcal{T}(X),\ZZ)
\end{equation*}
is an application of the well-known concept of $G$-vector
bundle as presented in~\cite{atiyah-k,segal}. We define an
action by $\mathcal{G}_0(X)$ on the trivial line bundle
$\tilde L =
\mathcal{P}(X) \times \CC$ by
\begin{equation}\label{g-action}
(\mu,z)\mapsto (\mu^g, C[\mu,g]z)\,.
\end{equation}
The action is free since it is so on the first factor, hence
$L=\tilde L/\mathcal{G}_0(X)$ is a line bundle over
$\mathcal{T}(X)$. As it is easily checked, cohomologous
cocycles yield isomorphic bundles, and so $L_{[C]}$ is
trivial if and only if $[C]$ is trivial.

Next, observe that $C[\mu,g]$ can be defined using the
quasi-Fuchsian prescription, which, according to
Theorem~\ref{theorem:c}, yields a holomorphic $W$. Moreover,
$\mu^g$ is holomorphic in $\mu$, as it follows from the
explicit expression. Thus, $C[\,\cdot\, ,g]$ is holomorphic
and so is the action~\eqref{g-action}. 
\end{proof}
\begin{remark}
The construction of the line bundle $L$ is well known from
works on anomalies~\cite{alv-g,catenacci,falqui}. An
explicit construction of the map $H^1(\mathcal{G}_0(X),
\CC^*(\mathcal{P}(X))) \rightarrow H^2(\mathcal{T}(X),\ZZ)$
using \v{C}ech cohomology appears in~\cite{falqui}.
\end{remark}
It follows from general arguments (cf.~\cite{segal}) that
sections of $L_{[C]}$ can be identified with the
$\mathcal{G}_0(X)$-invariant sections of $\tilde L$, namely
with those functions $\Phi:\mathcal{P}(X) \rightarrow \CC$
satisfying
\begin{equation*}
\Phi[\mu^g] = C[\mu,g]\, \Phi[\mu]\,.
\end{equation*}
Since the conformal block $\Psi=\exp(-W)$ does not vanish,
the foregoing proves the following 
\begin{proposition}
The conformal block $\Psi$ descends to a non-vanishing
section of $L_{[C]}$, thereby providing a trivializing
isomorphism $L_{[C]} \rightarrow \mathcal{T}(X)\times
\CC$.
\end{proposition}
Observe (cf. \cite{z2}) that the line bundle $L_{[C]}$ is
holomorphically trivial due to a general property of the
Teichm\"uller space being a contractible domain of
holomorphy~\cite{nag}. Our construction provides an instance
of this general fact, as well as an explicit trivializing
map. Also note that, due to the universal nature of the 
cocycle $C$, the ratio of two different conformal blocks,
in accordance with~\cite{ver}, is $\mathcal{G}_0(X)$-invariant
and, therefore, descends to a non-vanishing function on the 
Teichm\"{u}ller space $\mathcal{T}(X)$. 

\subsection{}
The preceding observations bring in several additional
questions concerning the geometrical significance of
$\exp(-W[\mu])$. For instance, we can define the trivial
connection on the trivial line bundle $\tilde L$ on
$\mathcal{P}(X)$: 
\begin{align*}
\nabla \Phi &= \Psi\d (\Psi^{-1}\Phi)\\
&= \d \Phi - (\Psi^{-1}\d\Psi )\Phi\,.
\end{align*}
This connection is easily verified to be
$\mathcal{G}_0(X)$-invariant, hence it descends onto
$L_{[C]}$. It follows from Proposition~\ref{restatement} and
Theorem~\ref{theorem:b} that the connection form coincides
with $\d W =c\,T/12\pi$.

This is very reminiscent of Friedan and Shenker's modular
geometry program for CFT~\cite{fs}, where the vacuum
expectation value of the stress-energy tensor is interpreted
as a connection on a line bundle over the moduli space. As
a further development, this suggests studying the action of
the full group $\mathcal{G}(X)$ on the presented
construction. As it is well known~\cite{ee}, the quotient of
$\mathcal{P}(X)/\mathcal{G}(X)$ (the action being the same
as in the previous case) is precisely the moduli space of
compact Riemann surfaces of genus $g>1$. All the local
formulas will stay the same, while the action of the modular
group $\mathcal{G}(X)/\mathcal{G}_0(X)$ on $\mathcal{T}(X)$
will introduce the topological ``twisting''. All of this
should be fundamental for the differential-geometrical
realization of Friedan and Shenker's program. In this
respect it is important, as we proved in the paper, that the
functional $W[\mu]$ is independent of the marking of a
Riemann surface $X$.

Another direction, more directly related to the Earle-Eells
fibration consists in finding the geometric interpretation of
the critical points $T=0$ and ``vertical critical'' points
$\mu_{zzz}=0$ of the functional $W[\mu]$.

Finally, the question of the relation of $W[\mu]$
with the full induced gravity action on $X$ is also very 
important. Recall the genus zero factorization~\cite{ver}
\begin{equation*}
\int R\Delta^{-1}R = W[\mu] + \overline{W}[\bar\mu] +
K[\phi,\mu,\bar\mu]\,, 
\end{equation*}
where the term $K[\phi,\mu,\bar\mu]$
is further decomposed as a sum
\begin{equation*}
K[\phi,\mu,\bar\mu] = S_L[\phi,\mu,\bar\mu] +
K_{\mathrm{BK}}[\mu,\bar\mu]
\end{equation*}
of the Belavin-Knizhnik-like anomaly term plus the Liouville
action in the background $\vert \d z +\mu\d\bar
z\vert^2$. After having properly defined $W[\mu]$ on $X$, it
is natural to ask whether such a decomposition holds in
higher genus as well. We observe that the general
(co)homological techniques applied in this paper can also be
used to give a mathematically rigorous construction of the
Liouville action (in various backgrounds) in the form of a
``bulk'' term plus boundary and vertex corrections, as in
the spirit of~\cite{t,zt}. A construction of this kind
should provide a meaning also to the full action $\int
R\Delta^{-1}R$ in terms of a Liouville action in the
``target'' complex structure, provided one can actually
define $K_\mathrm{BK}$ in higher genus as well. A full
understanding of the geometrical properties of $W[\mu]$ and
$K_\mathrm{BK}$ and their exponentials would be relevant in
order to put the Geometric Quantization approach of
ref.~\cite{ver} and, more generally, the three-dimensional
approach to two-dimensional gravity on a more conventional
mathematical basis. Finally, similar construction can be
carried out for defining the WZW functional on the higher
genus Riemann surfaces. We are planning to address these
questions in the next publications.
\appendix
\section{Some facts from Homological Algebra}\label{app:a}
We give a brief account on the use of double complexes as
applied to our situation.  We shall mainly focus on homology
and just indicate the required modifications to discuss the
cohomological counterpart of the various statements. For a
full account cf. any book on homological algebra, like, for
instance,~\cite{mac-lane}.
\subsection{} 
The framework we put ourselves in is sufficiently simple
that one can in fact avoid the use of spectral sequences
altogether in the proof of Lemmas~\ref{lemma-homology}
and~\ref{lemma-cohomology}, provided one takes into account
a few simple facts from homological algebra. The key point
is that the various double complexes we are interested in
have trivial (co)homology in higher degrees with respect to
either the first or second differentials, so the arguments
can be given in general, without referring to specific
examples.

Let $\KKK_{\bullet,\bullet}$ a double complex with
differentials $\delp:\KKK_{p,q}\rightarrow\KKK_{p-1,q}$ and
$\delpp:\KKK_{p,q}\rightarrow\KKK_{p,q-1}$, and total
differential $\del\bigr\rvert_{\KKK_{p,q}}=\delp
+(-1)^p\,\delpp$. According to our discussion, let us make
the assumption that
\begin{equation*}
H^{\delpp}_q(\KKK_{p,\bullet})=\begin{cases} C_p & q=0 \\ 0
			     & q>0\, .  \end{cases}
\end{equation*}
Then $\CCC_\bullet \eqdef\oplus C_p$ inherits a
differential\footnote{The use of the same symbol to denote
the differentials in $\CCC$ and $\Tot\KKK$ should not
generate any confusion.} $\del :C_p\rightarrow C_{p-1}$ from
the first differential $\delp$ in the double complex, and
since
\begin{equation*}
\cdots\overset{\delpp}{\longleftarrow}\KKK_{p,q-1}
\overset{\delpp}{\longleftarrow} \KKK_{p,q}
\overset{\delpp}{\longleftarrow}
\KKK_{p,q+1}\overset{\delpp}{\longleftarrow}\cdots
\end{equation*}
is exact except in degree zero, we can ``augment''
$\KKK_{\bullet,\bullet}$ inserting the projection
$\varepsilon :\KKK_{p,0}\rightarrow C_p$ to obtain the exact
sequence
\begin{equation*}
0\longleftarrow \CCC_\bullet \longleftarrow
\KKK_{\bullet,\bullet} \, .
\end{equation*}
\begin{proposition}\label{auxiliary}
\begin{equation*}
H_\bullet (\Tot\KKK )\cong H_\bullet (\CCC )\,.
\end{equation*}
\end{proposition}
\begin{proof}
This is a routine check of the definitions. Suppose $c\in
C_p$ is closed, i.e. $\del c=0$.  This means that it exists
a chain $c_0\in\KKK_{p,0}$ such that $\varepsilon (\delp
c_0)=0$, but $\varepsilon (\delp c_0)$ is the class
represented by $\delp c_0$, since we clearly have
$\delpp\delp c_0=0$. So, this class is zero, and therefore
we have
\begin{equation*}
\delp c_0 =\delpp c_1\quad \text{for}\; c_1\in\KKK_{p-1,1}\,
.
\end{equation*}
Now, $\delpp (\delp c_1) = \delp (\delpp c_1) = \delp\delp
c_0 =0$, and since the $\delpp$-homology of
$\KKK_{\bullet,\bullet}$ is concentrated only in dimension
zero, it must exist a $c_2\in\KKK_{p-2,2}$ such that
\begin{equation*}
\delp c_1 = \delpp c_2\, ,
\end{equation*}
and so on. The procedure stops at the $p$-th step. Thus the
chain
\begin{equation*}
C = c_0 + \sum_{i=1}^p\, (-1)^{\sum_{k=0}^{i-1}(p-k)}c_i
\end{equation*}
is a cycle in $\Tot\KKK$, that is, $\del C = 0$.

Conversely, suppose $C= c_0 + \sum_{i=1}^p\,
(-1)^{\sum_{k=0}^{i-1}(p-k)}c_i\in \Tot\KKK$ is
$\del$-closed. Then $c\equiv \varepsilon (c_0)$ is a degree
$p$ cycle in $\CCC_p$. Indeed, in degree $(p-1,0)$ we have
$\delp c_0=\delpp c_1$ and
\begin{equation*}
\varepsilon (\delp c_0)=\varepsilon (\delpp c_1)=0\, ,
\end{equation*}
since the augmentation is exact.

That the cycle $c\in C_p$ is a boundary if and only if
$C\in\Tot\KKK$ is a boundary can be proven along the same
lines. This completes the argument.
\end{proof}
\subsection{}
Recall from section~\ref{algebra} the various double
complexes we used. In particular, $\KKK_{\bullet,\bullet}
=\SSS_\bullet \otimes_{\ZZ\Gamma}\BBB_\bullet$ is the double
complex obtained tensoring the singular chain complex on
$X_0\cong\HH$ with the ``bar'' complex
\begin{equation}\label{bar}
0\longleftarrow\BBB_0 \overset{\delpp}{\longleftarrow}\BBB_1
\overset{\delpp}{\longleftarrow}\cdots
\overset{\delpp}{\longleftarrow}\BBB_n
\overset{\delpp}{\longleftarrow}\cdots
\end{equation}
which is exact except in degree zero. Its definition has
been given in the main text. Being $\BBB_0$ a
$\Gamma$-module on the generator $[\; ]$, introducing the
augmentation $\varepsilon:\BBB_0\rightarrow\ZZ$,
$\varepsilon ([\; ])=1$, we can rewrite it as the exact
sequence
\begin{equation}\label{augmented-bar}
0\longleftarrow\ZZ
\overset{\varepsilon}{\longleftarrow}\BBB_0
\overset{\delpp}{\longleftarrow}\BBB_1
\overset{\delpp}{\longleftarrow}\cdots
\overset{\delpp}{\longleftarrow}\BBB_n
\overset{\delpp}{\longleftarrow}\cdots\, .
\end{equation}
The above exact sequence is usually referred to as a
``resolution'' of the integers. Since every $\BBB_q$ is a
\emph{free} $\Gamma$-module, the sequence is a free
resolution.

The singular chain complex $\SSS_\bullet \equiv S_\bullet
(X_0)$ needs little description. Since $\Gamma$ acts on the
space, $\SSS_\bullet$ acquires a $\Gamma$-module structure
simply by translating around the chains. That this actually
is a complex of \emph{free} $\Gamma$-modules is proven
in~\cite{mac-lane} or~\cite{brown}. A choice of free
generators is to take those chains whose first vertex lies
in a suitably chosen fundamental domain in $X_0$. The
differential, which we called $\delp$ in the main text, is
just the usual boundary homomorphism.

The homology of $\Gamma$ with coefficients in any
$\Gamma$-module $M$ is by definition the homology of the
complex $M\otimes_{\ZZ\Gamma}\BBB_\bullet$.  (Any other
resolution of \ZZ\ would be adequate.) In fact, tensor
product does not preserve exactness in general. As a matter
of terminology, a module $M$ such that any exact sequence
remains exact after tensoring with it, is called
\emph{flat}. Therefore, all the higher homology groups of
$\Gamma$ with coefficient in a flat module will be zero. A
free $\Gamma$-module is in particular flat, as it is very
easy to see. So, in our case, we have
\begin{equation*}
H_q(\Gamma ,\SSS_p) = \begin{cases}
		      \SSS_p\otimes_{\ZZ\Gamma}\ZZ& q=0\\ 0&
		      q > 0 \end{cases}
\end{equation*}
where \ZZ\ is considered as a trivial
$\Gamma$-module. Moreover, note that $\SSS_p
\otimes_{\ZZ\Gamma}\ZZ \equiv S_p(X_0)\otimes_{\ZZ\Gamma}\ZZ
\cong S_p(X)$ the space of singular chains on the surface.
Indeed, if $c$ is any chain on $X_0$ and $\gamma$ is any
group element, we have $c\cdot\gamma\otimes 1=c\otimes
\gamma\cdot 1 =c\otimes 1$, and therefore $c\otimes 1$ can
be identified with a singular chain on the surface, as
claimed.

After this preparations, we can exploit the exact
complex~\eqref{augmented-bar} to build the augmented double
complex
\begin{equation}\label{augmented-double}
\begin{CD}
0 @<<< \SSS_\bullet\otimes_{\ZZ\Gamma}\ZZ
@<{\id\otimes\varepsilon}<<
\SSS_\bullet\otimes_{\ZZ\Gamma}\BBB_\bullet
\end{CD}
\end{equation}
with exact rows. According to the foregoing, the leftmost
column in~\eqref{augmented-double} is to be identified with
the singular chain complex on the surface. (Or, more
generally, of the quotient space.)

The complex~\eqref{augmented-double} satisfies the
hypotheses of Proposition~\ref{auxiliary}, and since the
group homology is the $\delpp$-homology of the double
complex, we conclude that $H_\bullet (\Tot\KKK )\cong
H_\bullet (X,\ZZ )$ thereby proving one half of
Lemma~\ref{lemma-homology}.

In order to prove the other half, let us observe that
actually all the columns in~\ref{augmented-double}, except
the first one, are exact, $X_0\cong\HH$ being a contractible
space. Indeed, the complex $\SSS_\bullet$ carries no
homology except in degree zero, and we can ``augment'' it as
well to obtain another resolution of the integers:
\begin{equation*}
0\longleftarrow\ZZ
\overset{\varepsilon}{\longleftarrow}\SSS_0
\overset{\delp}{\longleftarrow}\SSS_1
\overset{\delp}{\longleftarrow}\cdots
\overset{\delp}{\longleftarrow}\SSS_n
\overset{\delp}{\longleftarrow}\cdots\, .
\end{equation*}
Now the situation is completely symmetric and we can just
``transpose'' the above constructions to build the augmented
complex
\begin{equation*}
\begin{CD}
\SSS_\bullet\otimes_{\ZZ\Gamma}\BBB_\bullet \\
@VV{\varepsilon\otimes\id}V \\
\ZZ\otimes_{\ZZ\Gamma}\BBB_\bullet\\ @VVV\\0
\end{CD}
\end{equation*}
and apply Proposition~\ref{auxiliary} to it to show that
$H_\bullet (\Tot\KKK )\cong H_\bullet (\Gamma, \ZZ )$.
\subsection{} 
The cohomological picture has a very similar structure. The
cohomology of $\Gamma$ with coefficients in $M$ is by
definition the homology of the complex
$\Hom_{\ZZ\Gamma}(\BBB_\bullet ,M)$. (Notice that $\Hom$ is
contravariant in the first variable, thus it reverses the
arrows.) We will be in position to apply the analogous of
Proposition \ref{auxiliary} with the arrows reversed to the
complex $\CCC^{\bullet,\bullet} =\Hom (\BBB_\bullet
,\AAA^\bullet)$ provided we show that $H^q(\Gamma
,\AAA^p)=0$ for $q>0$, that is, $\Hom(\,\cdot\, ,\AAA^p)$
must preserve exactness, so that the higher cohomology
groups are zero.  An \emph{injective} module $M$ is by
definition a $\Gamma$-module such that $\Hom (\,\cdot\, ,M)$
preserves exactness, hence the higher cohomology groups of
$\Gamma$ with coefficients into an injective are zero. Thus
we have to show that $\AAA^p$ is injective as a
$\Gamma$-module. In fact, more can be done, namely it can be
shown that $\AAA^p\cong \Hom_{\ZZ}(\ZZ\Gamma ,A^p_\CC(X))$,
where $A^p_\CC (X)$ is the vector space of (complex valued)
differential forms on the Riemann surface $X$. The (easy)
proof of this assertion requires the construction of an
equivariant partition of unity on $\HH$,
see~\cite{kra}. Then $\AAA^p$ has no higher cohomology since
\begin{equation*}
\begin{split}
\Hom_{\ZZ\Gamma}(\BBB_\bullet,\AAA^p) &\cong
\Hom_{\ZZ\Gamma}(\BBB_\bullet,\Hom_\ZZ(\ZZ\Gamma
,A^p_\CC(X)))\\ &\cong \Hom_\ZZ (\BBB_\bullet,A^p_\CC(X))\,
,
\end{split}
\end{equation*}
and the last complex has no cohomology, except in degree
zero. Thus we have
\begin{equation*}
H^q(\Gamma,\AAA^p)=\begin{cases} A^p_\CC(X)& q=0\\ 0& q >
		   0\, , \end{cases}
\end{equation*}
and applying Proposition~\ref{auxiliary} to the double
complex $\CCC^{\bullet,\bullet}$ we can prove that
\begin{equation*}
H^\bullet (\Tot\CCC )\cong H^\bullet (X,\CC )\, .
\end{equation*}
To prove the rest of Lemma~\ref{lemma-cohomology} we need
only use the contractibility of $X_0\cong\HH$, so that
$\AAA^\bullet$ has no cohomology, and apply
Proposition~\ref{auxiliary} to the transposed double
complex.
\subsection*{Acknowledgements} We would like to thank
J. L. Dupont, C.-H. Sah and S. Shatashvili for very helpful
discussions. We also thank R. Zucchini and G. Falqui for
kindly pointing out several references to previous works on
the subject.\\ The work of E.A. was supported by the
National Research Council (CNR), Italy; the work of L.T. was
partially supported by the NSF grant DMS-95-00557.

\end{document}